\documentclass[aps,prd,twocolumn,nofootinbib,amsmath,amssymb,showpacs]{revtex4-1}

\usepackage{graphics}
\usepackage[usenames,dvipsnames]{color}
\usepackage{xcolor}
\usepackage{mathtools}
\usepackage{hyperref}
\hypersetup{
    colorlinks=true,
        linkcolor={blue!80!black},
        citecolor={blue!80!black},
        urlcolor={blue!80!black} 
}
\usepackage{mwe}
\usepackage[subnum]{cases}
\usepackage{lipsum}

\begin{document}
 
\title{Radial oscillations in neutron stars from QCD}

\author{Jos\'e C. {\sc Jim\'enez$^1$}}
\author{Eduardo S. {\sc Fraga$^2$}}

\affiliation{$^1$Instituto de F\'isica, Universidade de S\~ao Paulo, \\ Rua do Mat\~ao, 1371, Butant\~a,  05508-090, S\~ao Paulo, Brazil\\
$^2$Instituto de F\'\i sica, Universidade Federal do Rio de Janeiro,
Caixa Postal 68528, 21941-972, Rio de Janeiro, RJ, Brazil}

\date{\today}


\begin{abstract}
We study the stability against infinitesimal radial oscillations of neutron stars generated by a set of equations of state obtained from first-principle calculations in cold and dense QCD and constrained by observational data. We consider mild and large violations of the conformal bound, $c_{s}=1/\sqrt{3}$, in stars that can possibly contain a quark matter core. Some neutron star families in the mass-radius diagram become dynamically unstable due to large oscillation amplitudes near the core.
\end{abstract}

\pacs{04.40.Dg,97.60.Jd,12.38.Mh,25.75.Nq}


\maketitle

\section{Introduction}
\label{sec-intro}

The equation of state (EoS) for neutron star matter, including the constraints from electric charge neutrality and beta equilibrium,
is needed as the input to determine the structural properties of neutron stars (NS). It has to cover a wide range of densities, from below the nuclear saturation density, $n_{0}=0.16{~\rm fm^{-3}}$, up to about $10n_{0}$ \cite{Glendenning:2000}, a real challenge. At such high densities, the question whether quark matter (QM) would be present in the core of NS arises naturally (see, e.g., Refs. \cite{Lattimer:2004pg,Alford:2006vz}). Some possibilities for its formation mechanism being core-collapse supernova explosions \cite{Sagert:2008ka} and the merger of relatively light NS \cite{Weih:2019xvw}. Ultimately, this question must be settled by astrophysical observations. For instance, the NICER mission has recently shown the radii of NS with masses $M\sim 1.4 M_{\odot}$ to be $R\sim 13$km  with a very good precision \cite{Miller:2019cac,Riley:2019yda}. Still, one expects to find QM only in the interior of very massive NS, $2 M_{\odot}\lesssim{M}\lesssim 2.2 M_{\odot}$ \cite{Demorest:2010bx,Fonseca:2016tux,Antoniadis:2013pzd,Cromartie:2019kug,Linares:2018ppq}, where central densities could be high enough. 

In order to build the EoS for neutron star matter starting from first principles and perform systematic, controlled approximations, one can combine chiral effective field theory for nuclear matter at low baryon densities ($n_{B}\lesssim{n}_{0}$) \cite{Tews:2012fj,Hebeler:2013nza} with perturbative QCD (pQCD) for deconfined QM at very high densities ($n_{B}\gg{n}_{0}$) \cite{Kurkela:2009gj,Fraga:2013qra}. The non-trivial sector of intermediate densities can be parametrized by a polytropic interpolation taking into account the two-solar mass constraint \cite{Kurkela:2014vha} and also the gravitational wave event GW170817 \cite{Annala:2017llu}. The maximum mass of NS can also be constrained to be in the range $2.1 M_{\odot}\lesssim{M}\lesssim 2.3 M_{\odot}$ \cite{Rezzolla:2017aly,Ruiz:2017due}.

The EoS for dilute nuclear matter, $n_B {\lesssim} n_{0}$, was obtained using the ab initio approach of chiral effective field theory for light nucleons up to next-to-next-to-next-to-leading order in chiral perturbation theory ($\chi$PT) by Tews {\it et al.} \cite{Tews:2012fj}. To describe neutron star physics, $\beta$ equilibrium and electric charge neutrality were implemented by Hebeler {\it et al.} \cite{Hebeler:2013nza}. The authors, then, extrapolate to densities above $n_{B}=1.1n_{0}$ by performing  successive matchings onto polytropic equations of state. 

Following Hebeler {\it et al.}, Kurkela {\it et al.} \cite{Kurkela:2014vha} added the perturbative QCD description at very high densities, being the polytropes used now as interpolations. In doing so, the extremes of low and high densities could be described from first principles, constraining the behavior of the neutron star equation of state. The pressure at high density had been computed to three loops in Ref. \cite{Kurkela:2009gj}, including renormalization group effects for the strange quark mass and strong coupling $\alpha_{s}$ up to $\mathcal{O}(\alpha^{2}_{s})$. By adding the external constraint from the measurements of NS with masses around $2M_{\odot}$ \cite{Demorest:2010bx,Antoniadis:2013pzd}, the relative uncertainty of the EoS band is reduced to less than $30\%$ at all relevant densities. Limits on the tidal deformability, $\Lambda$, provided by the binary merger GW170817 \cite{TheLIGOScientific:2017qsa} restrict even more the EoS \cite{Annala:2017llu}.

Recently, Annala \textit{et al.} \cite{Annala:2019puf} allowed large deviations of the adiabatic squared speed of sound, $c^{2}_{\rm s}=dP/d\epsilon$, from the conformal bound of $1/3$ going up to the causality limit of $1$, when performing the matching of EoSs through their new speed-of-sound interpolation method (see also Ref. \cite{Tews:2018iwm}). Their uncertainty bands display a rapid softening from stiff nuclear matter to soft quark matter, i.e. their modified polytropic index\footnote{Defined as $\gamma\equiv{d(\log{P})/d(\log{\epsilon})}$ and being related to the usual adiabatic index $\Gamma\equiv{d(\log{P})/d(\log{n})}$ by $\Gamma=\gamma+c^{2}_{s}$.}, $\gamma$, changes from $\gamma_{\rm nucl}\gtrsim{2.5}$ to almost conformal QCD matter, $\gamma_{\rm pQCD}\sim{1}$, at  $\epsilon_{\rm crit}~{\approx}~$400--700 MeV$/{\rm fm}^{3}$. These are typical energy densities for the deconfinement transition at high temperatures found in ultrarelativistic heavy ion collisions \cite{Bazavov:2014pvz}.  

\begin{figure*}[t!]
\begin{center}
\hbox{
\includegraphics[width=0.495\textwidth]{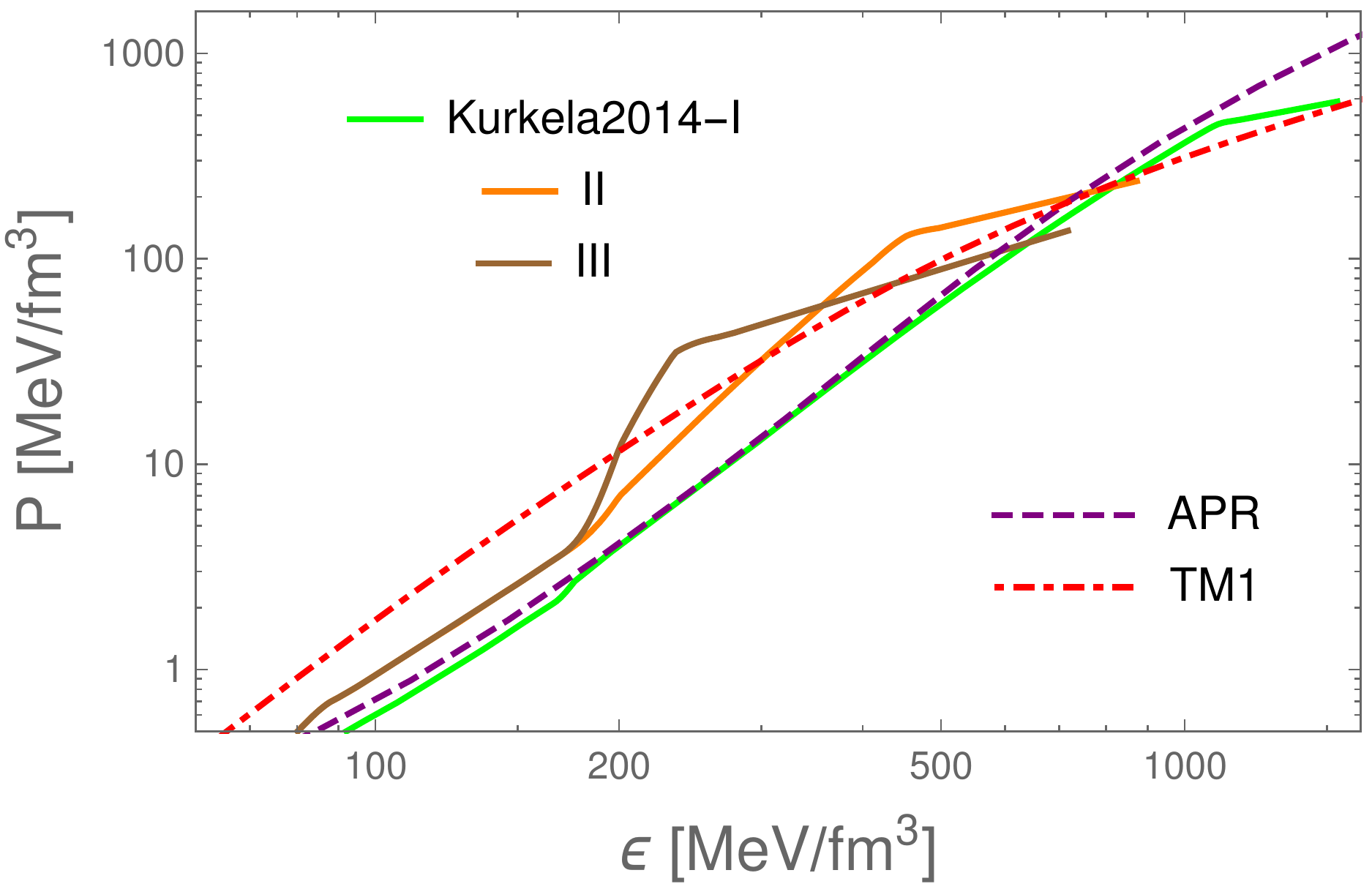} 	\includegraphics[width=0.495\textwidth]{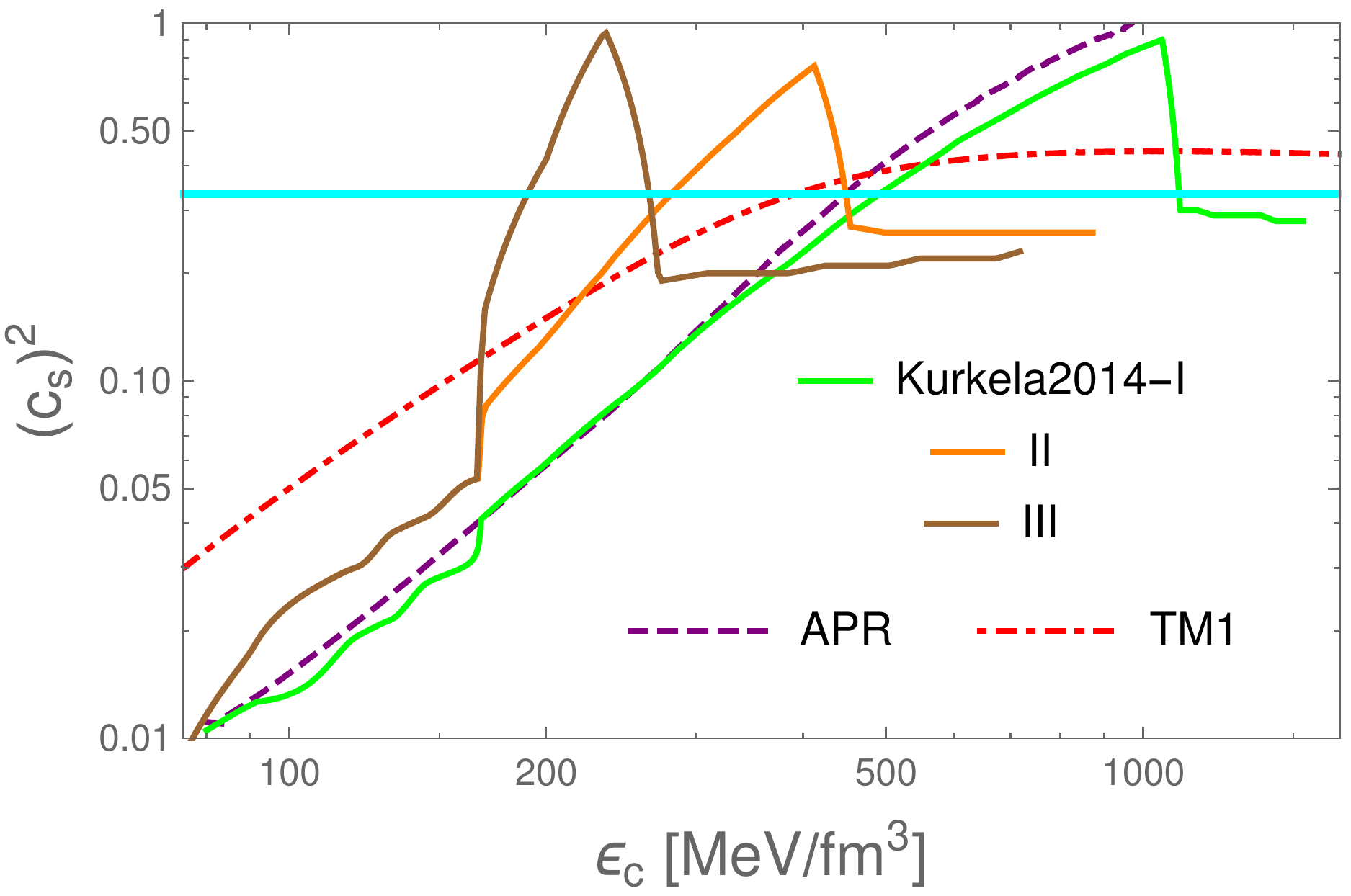}}
\caption{Left panel: Set of EoSs from Ref. \cite{Kurkela:2014vha} (I, II and III) analyzed in this work. Right panel: Respective squared speed of sound ($c^{2}_{s}$) vs central energy density. The horizontal cyan line represents the conformal limit $c^{2}_{s}=1/3$. For comparison we also show results from APR \cite{Akmal:1998cf} and TM1 \cite{Shen:1998gq} equations of state for pure nuclear matter.}
\label{fig:EoSsKurk}
\end{center}
\end{figure*}

By means of a careful analysis of the sudden bending of their polytropic index, Annala \textit{et al.} \cite{Annala:2019puf} suggested that cold QM might exist in the core of the most massive NS, their presence being seizeable when $c^{2}_{s,\rm max}<{0.5}$ and independent of the nature of the transition, whereas one would obtain pure hadronic stars only if $c^{2}_{s,\rm max}>{0.7}$ and one has a first-order transition (or a rapid crossover). The authors argued that the presence of QM at the core of heavy NS should be considered {\it the standard scenario and not an exotic alternative}. In particular, they find that, if the conformal limit is not strongly broken, i.e. $c^{2}_{s}<0.4$, a significant amount of QM might exist in massive NS.

In this paper we investigate the dynamical stability against infinitesimal radial pulsations of NS generated by a set of EoSs extracted from the constrained bands considered in Ref. \cite{Annala:2019puf}. We use the framework of Gondek \textit{et al.}  \cite{Gondek:1997fd} to compute the fundamental mode frequencies ($f_{n=0}$) for each EoS. We use these results to show novel features of stable NS with (without) QM cores having sub-conformal (extreme) values of $c^{2}_{s}$, in terms of the adiabatic index $\Gamma$. We adopt three representative (tabulated) EoS from Ref. \cite{Annala:2019puf}, producing heavy NS satisfying multi-messenger bounds. They correspond to a sub-conformal EoS with a crossover transition leading to a seizeable QM core (Case I), an intermediate-$c^{2}_{s}$ EoS with a crossover transition leading also to a seizeable QM core (Case II), and a high-$c^{2}_{s}$ EoS with a strong first-order phase transition without a QM core (Case III). 

\begin{figure*}[t!]
\begin{center}
\hbox{
\includegraphics[width=0.495\textwidth]{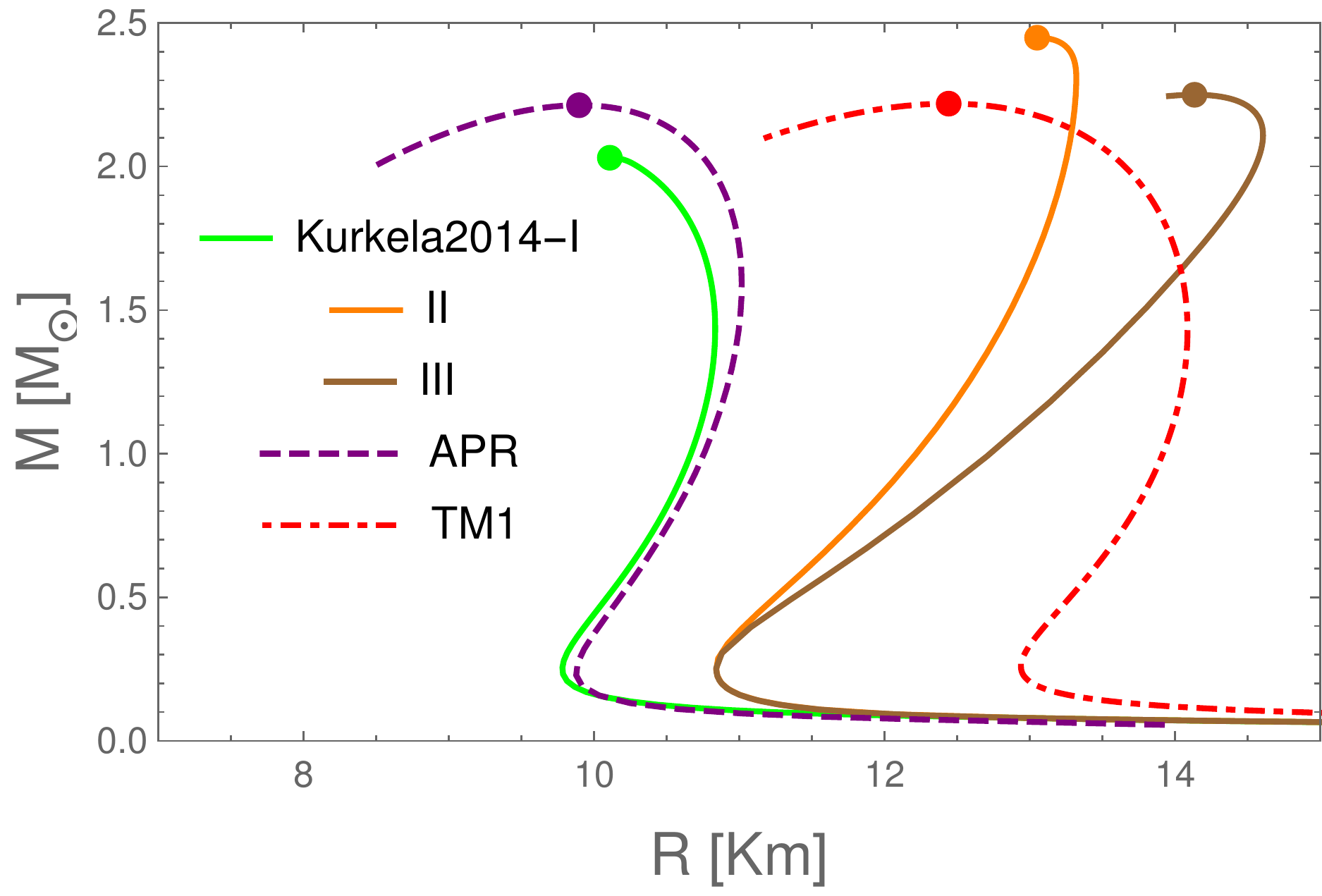} 	\includegraphics[width=0.495\textwidth]{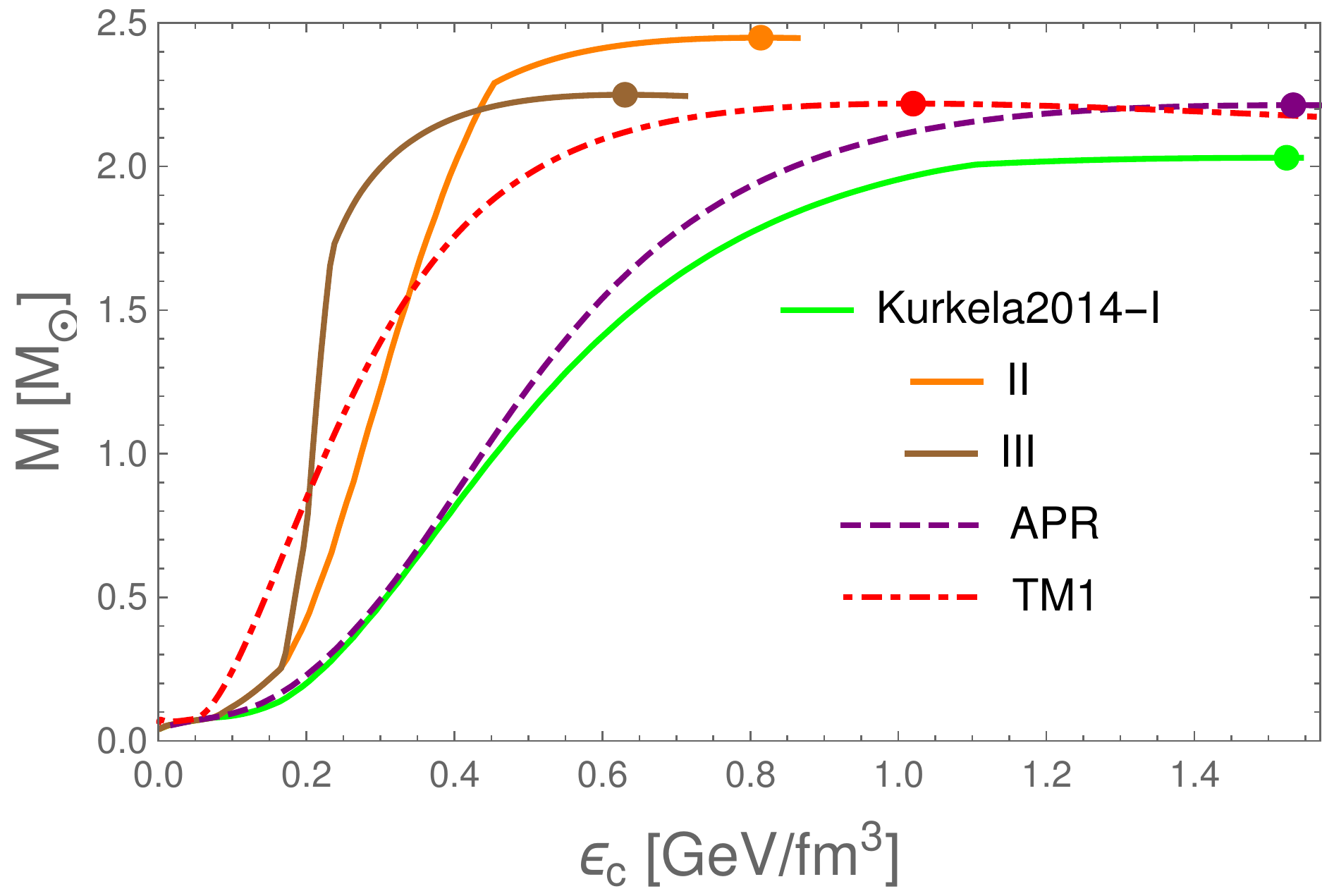}}
\hbox{
\includegraphics[width=0.495\textwidth]{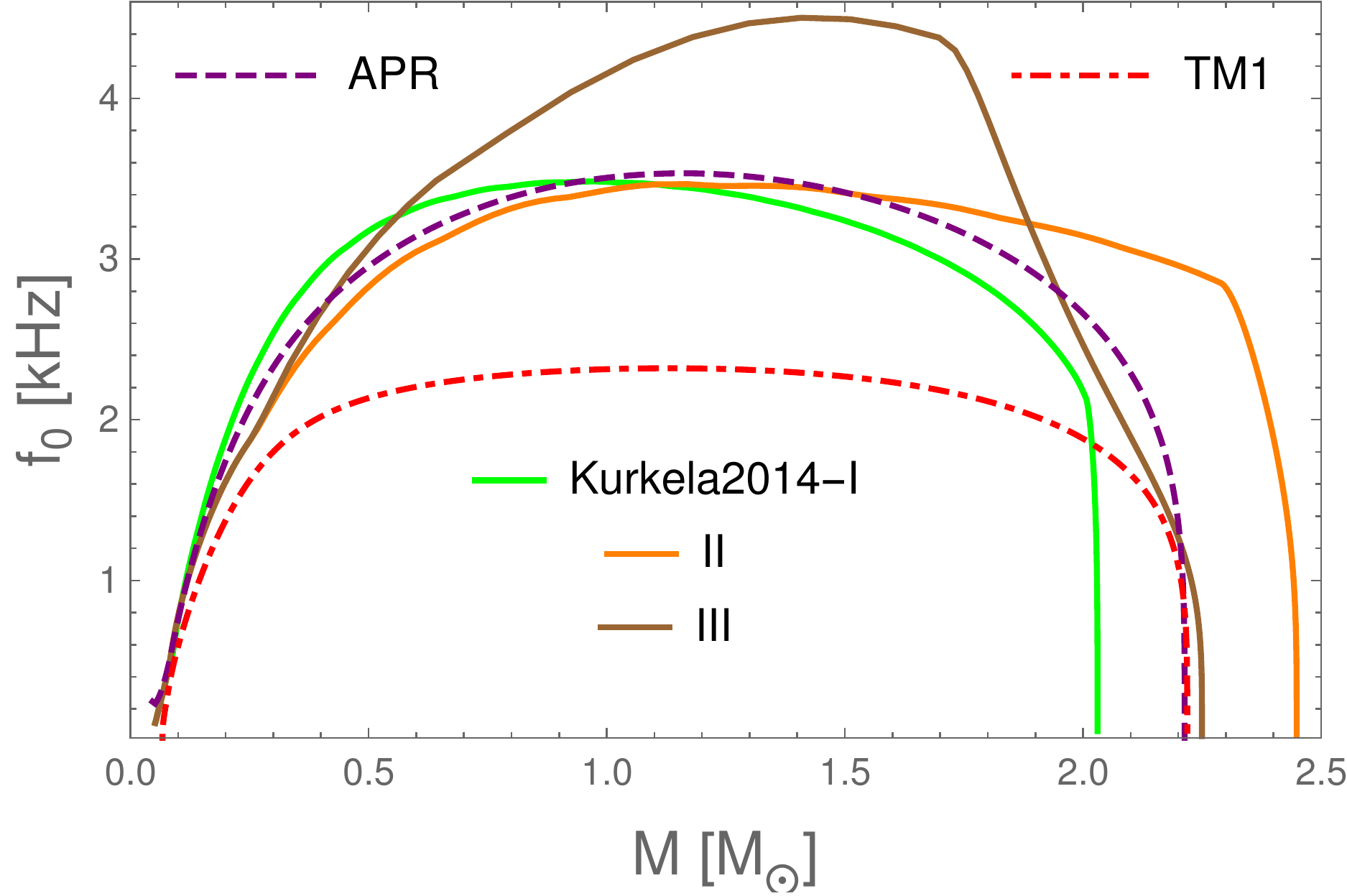} \includegraphics[width=0.495\textwidth]{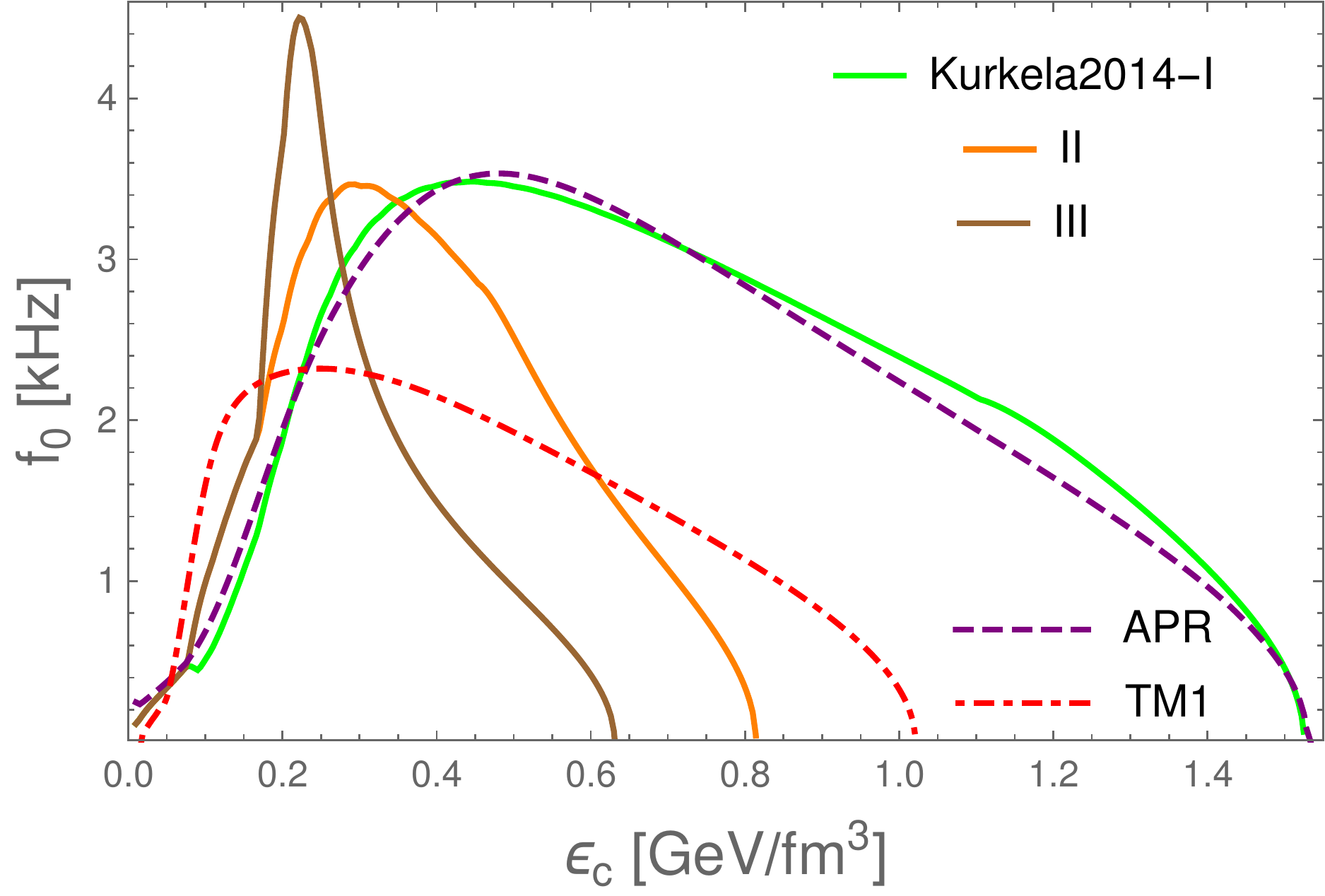}}
\caption{Upper panels: mass-radius diagram and mass as a function of central energy density obtained for the EoS of Kurkela {\it et al.} \cite{Kurkela:2014vha}. Dots indicate the end of the star sequence at $M_{\rm max}$, i.e. $\partial{M}/\partial{\epsilon_{c}}=0$. Lower panels: corresponding fundamental-mode frequencies ($f_{0}$) as functions of the gravitational mass and central energy density. For comparison we show results from APR \cite{Akmal:1998cf} and TM1 \cite{Shen:1998gq} equations of state for pure nuclear matter.}
\label{fig:Kurk2014}
\end{center}
\end{figure*}

We also study three other EoSs from constrained bands \cite{Gorda:2019abc} satisfying $c^{2}_{\rm max, s}>0.7$ with first-order transitions (Cases 1, 3, 4) and one having a crossover transition with $c^{2}_{s}<0.5$ (Case 2). The large $c^{2}_{s}$ in the former set implies a decreasing number of degrees of freedom at increasing densities \cite{Bedaque:2014sqa}, something which is in sharp contrast with hot QCD where quarks and gluons are liberated at high temperatures \cite{Gardim:2019xjs}. A partial resolution of this issue was given in Ref. \cite{Annala:2019puf}, where the authors suggested that NS with $c^{2}_{s}>0.4$ could be dismissed if one includes further astrophysical constraints, e.g. the electromagnetic counterpart of GW170817 \cite{Radice:2018ozg}. However, the unusual lumps at intermediate NS masses on the right-hand side of their corresponding mass-radius diagram \cite{Annala:2019puf} cannot be discarded using only observational constraints. 

Here, we use Cases 3 and 4 (without loss of generality) to show that those prominent lumps are dynamically unstable, although not in the usual sense of having the (squared) fundamental-mode frequency negative, i.e. $f^{2}_{n=0}<0$. Instead, their respective infinitesimal Lagrangian radial displacement, $\xi_{n=0}=(\Delta{r}/r)_{n=0}$, produces large values ($\sim{30}$), in contrast to the condition of being small, i.e. $0<\xi\leqslant{1}$, thereby allowing us to classify them, strictly speaking, as metastable. In fact, since radial oscillations with finite amplitudes induce non-linear effects that dynamically destabilize the star \cite{Gourgoulhon:1995fde}, one can exclude such families of NS from the interpolation scheme.
 
\begin{figure*}[t!]
\begin{center}
\hbox{\includegraphics[width=0.5\textwidth]{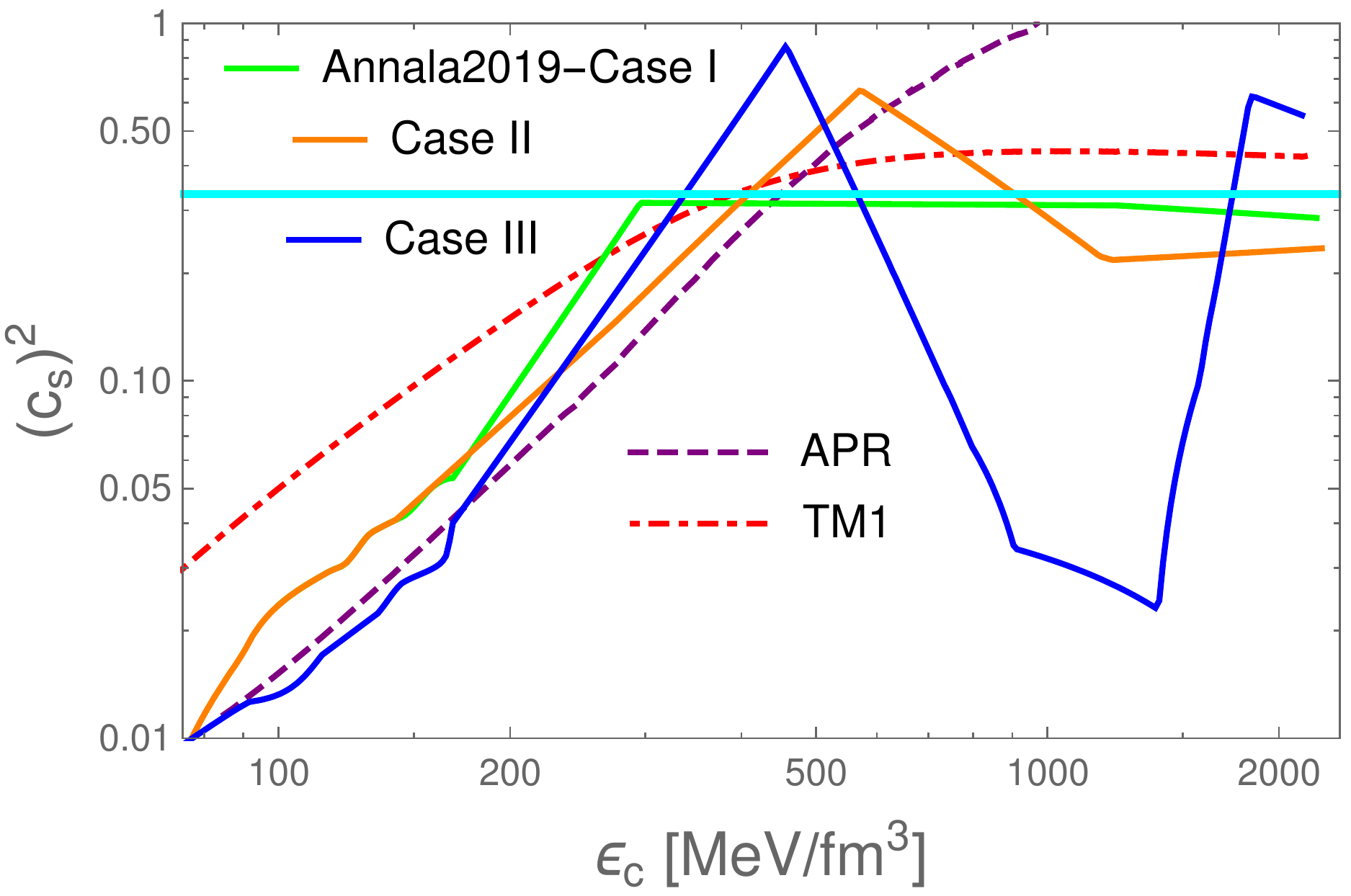}		\includegraphics[width=0.5\textwidth]{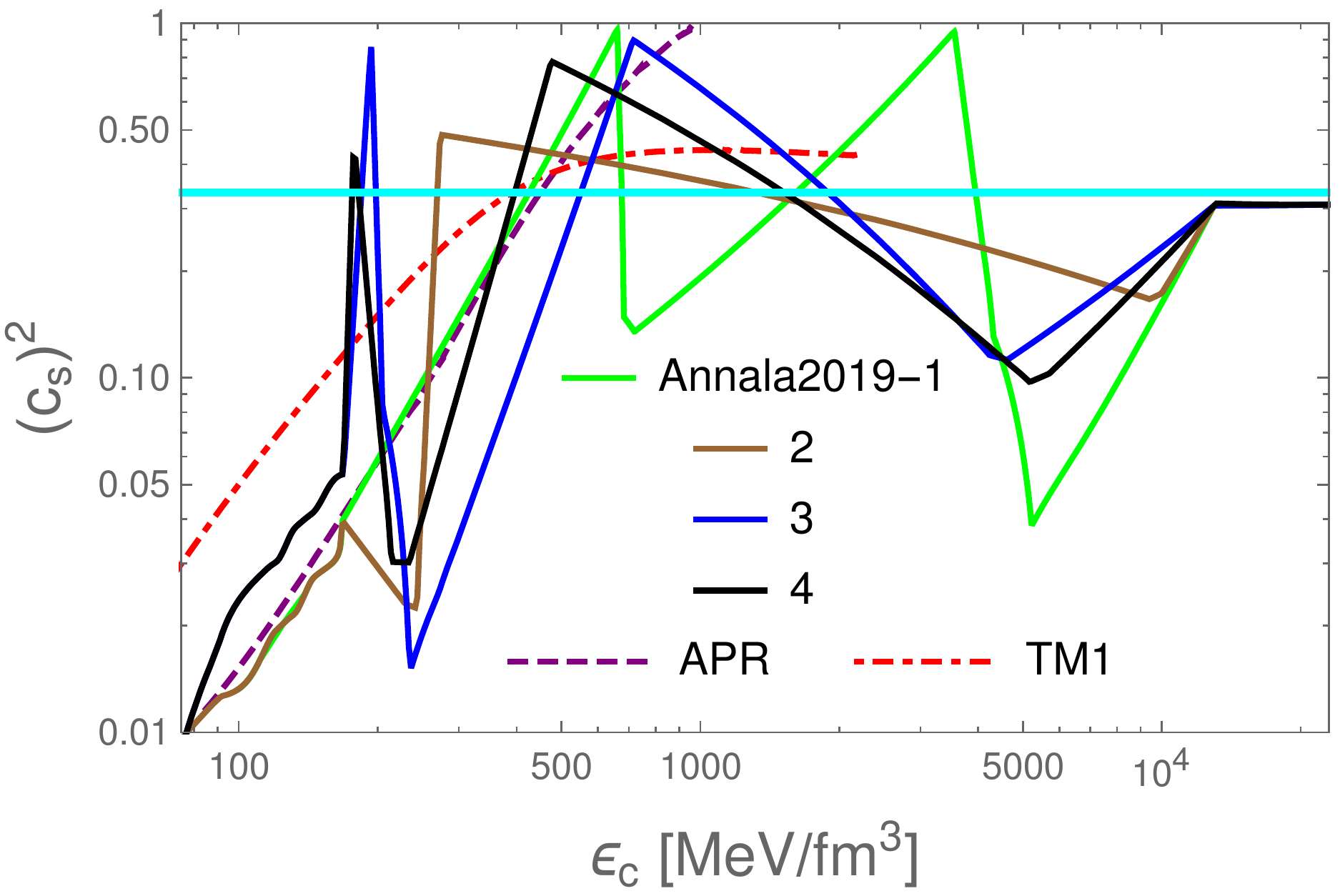}}
\hbox{
\includegraphics[width=0.495\textwidth]{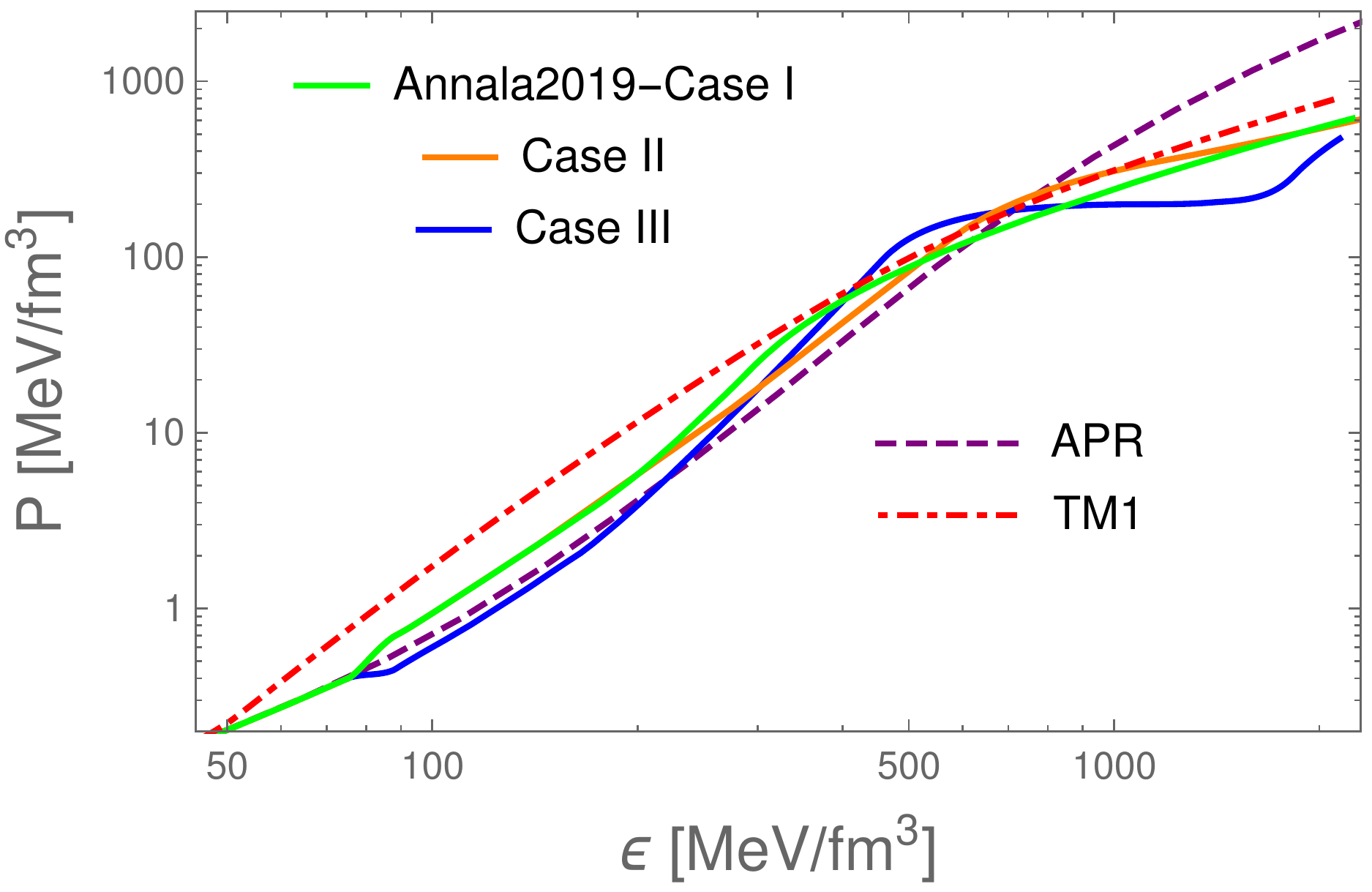} \includegraphics[width=0.495\textwidth]{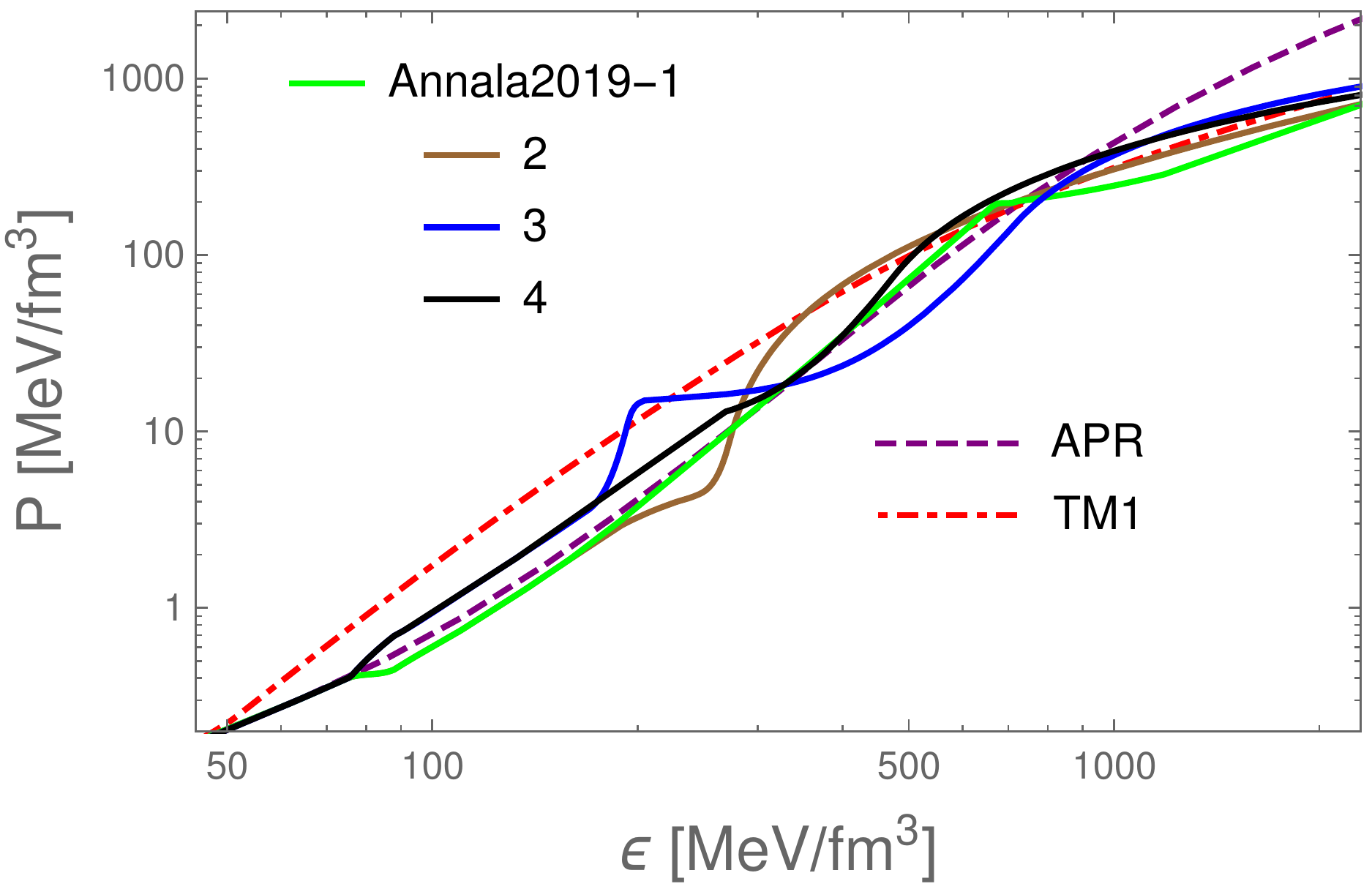}}
\caption{Upper panels: Squared speed-of-sound ($c^{2}_{s}$) vs central energy density for the constrained EoSs of Annala {\it et al.} \cite{Annala:2019puf} (Cases I, II and III; left panel) and additional EoSs \cite{Gorda:2019abc} producing the odd lump in the mass-radius diagram (Cases 1,2,3,4; right panel). Lower panels: Corresponding EoSs.}
\label{fig:EoSs}
\end{center}
\end{figure*}

For completeness, we also compare our results to the ones obtained from a pair of standard pure nuclear matter EoSs: APR \cite{Akmal:1998cf}, calculated using many-body techniques with phenomenological potentials,  and TM1 \cite{Shen:1998gq}, obtained within relativistic mean field theory. The form of these EoSs is depicted in Fig. \ref{fig:EoSs}.

The paper is organized as follows. In Sec. \ref{sec:stability} we summarize the main features of the first-order radial stability equations used in this paper. Section \ref{sec:results} shows our results for the behavior of the fundamental mode frequencies as a function of different stellar properties. Section \ref{sec:conclusion} presents our summary.

\section{Stability of the Fundamental Radial Mode}
\label{sec:stability}

The relation between the adiabatic index $\Gamma$ (via $c^{2}_{s}$) and the existence of QM in the core of NS brings the question of their dynamical stability against linear (small) radial perturbations, where $\Gamma$ plays a fundamental role. Imposing the condition $\partial{M}/\partial{\epsilon_{c}}\geq0$ plus astrophysical constraints helps reducing the band of EoS possibilities and, therefore, also the allowed region in the mass-radius diagram. However, to ensure full dynamical stability one needs to analyze radial pulsations through the behavior of the fundamental mode frequencies, $f_{n=0}$, which may be seen as a further constraint imposed by full general relativity.    

Radial pulsation in NS with QM cores were studied using simple models for the deconfined phase, e.g. the MIT bag model, and sometimes assuming a (dis)continuous transition. For instance, Gupta \textit{et al.} \cite{Gupta:2002fk} considered a Glendenning construction \cite{Glendenning:2000} for the mixed phase, i.e. a continuous transition (see Ref. \cite{Sahu:2001iv} for similar studies). This allows them to use the Sturm-Liouville approach originally derived for non-hybrid NS by Chandrashekar \cite{Chandrasekhar:1964zza}. They found a reduction in the maximum mass ($\Delta M_{\rm max} \sim 0.3 M_{\odot}$) and a new ``kink'' in the $\omega_{n}$ vs $\epsilon_{c}$ plane at the start of the mixed phase. On the other hand, first-order transitions might also occur in the interior of NS, leading to a more complicated analysis that requires the solution of supplementary equations \cite{Haensel:1989wax,Pereira:2017rmp}. The constrained neutron star matter EoSs of Annala \textit{et al.} \cite{Annala:2019puf}, on the other hand, exhibit first-order transitions that behave numerically as continuous rapid crossovers at the borders of the approximately flat mixed phase region. So, we can adopt the formalism of Gondek \textit{et al.} \cite{Gondek:1997fd}, which is mathematically equivalent to the one developed by Chandrashekar and has the advantage of being more amenable to numerical calculations: one can easily impose the boundary conditions at the surface of the star and there is no need for derivatives of the adiabatic index $\Gamma$ (see Appendix).  

The formalism consists of a pair of coupled first-order differential equations for the relative radial displacement $\Delta{r}/r\equiv\xi$ and Lagrangian perturbed pressure $\Delta{P}$. They can be written in matrix form as \cite{Gondek:1997fd}:
	\begin{equation}
  \begin{pmatrix}
   \dfrac{\strut d\Delta{P}}{\strut dr} \vspace{0.2cm}
    \\
    \dfrac{\strut d\xi}{\strut dr}
  \end{pmatrix}
  =
  \begin{pmatrix}
    \hspace{0.1cm}\mathcal{Z}(r) & \hspace{0.2cm} \mathcal{Q}(r, \omega^{2}) \vspace{0.9cm}
    \\
    \hspace{0.1cm}\mathcal{R}(r)\vspace{0.2cm} & \hspace{0.2cm} \mathcal{S}(r)
  \end{pmatrix}
  \begin{pmatrix}
    \Delta{P} \vspace{0.9cm}
    \\
    \vspace{0.3cm}\xi
  \end{pmatrix} \;,
  \label{Eq.Gondek}
\end{equation}
where $\omega$ is the oscillation frequency and $\mathcal{Z}$, $\mathcal{Q}$, $\mathcal{R}$, $\mathcal{S}$ are complicated functions of the radial profiles for the pressure $P(r)$, energy densitie $\epsilon (r)$, $\Gamma(r)$, metric functions $\nu(r)$ and $\lambda(r)$, after solving the Tolman-Oppenheimer-Volkov (TOV) equations for each EoS (for more details, see Appendix).

Dynamically stable NS are usually only characterized by the behavior of the lowest eigenvalue (fundamental mode), $\omega^{2}_{n=0}=(2\pi{f}_{n=0})^{2}$, $f_{n=0}$ being the (zero-mode) linear frequency which we use for numerical convenience. If $f^{2}_{n=0}>0$, then all $f^{2}_{n}>0$ and the star is stable. If $f^{2}_{n=0}<0$, then there is at least one unstable mode and the star becomes unstable. The transition between these two stellar states occurs when ${f}_{0}\rightarrow{0}$ \cite{Haensel:2007}.

\section{Results}
  \label{sec:results}

By solving the oscillation equations for the representative tabulated EoSs discussed in Section \ref{sec-intro}, one can obtain the frequencies for the oscillation modes $n=0,1 \dots.$, $f_{0}$ being the most relevant for the stability analysis. 

Here, we compute the frequency dependence on the total gravitational mass, $M$, and stellar central energy density, $\epsilon_{c}$. Besides the cases presented in Ref. \cite{Annala:2019puf}, we also consider, for comparison, the three representative bitropic EoSs (Cases I, II, III) of Ref. \cite{Kurkela:2014vha}. As shown in Fig. \ref{fig:EoSsKurk}, they surpass the bound $c^{2}_{s}=1/3$ (without a 1st-order transition) between monotropes rapidly in Cases I and III and slowly in Case II. As expected, all these features are manifested in their stellar structure as well as in the planes $f_{0}$ vs $M$ and $f_{0}$ vs $\epsilon_{c}$, as is clear in Fig. \ref{fig:Kurk2014}. In this figure, and analogous figures in what follows, one can see that static, $\partial{M}/\partial{\epsilon_{c}}=0$, and dynamical, $f_{0} \to 0$, stability criteria coincide at $M_{\rm max}$.

\begin{figure*}[t!]
\begin{center}
\hbox{
\includegraphics[width=0.495\textwidth]{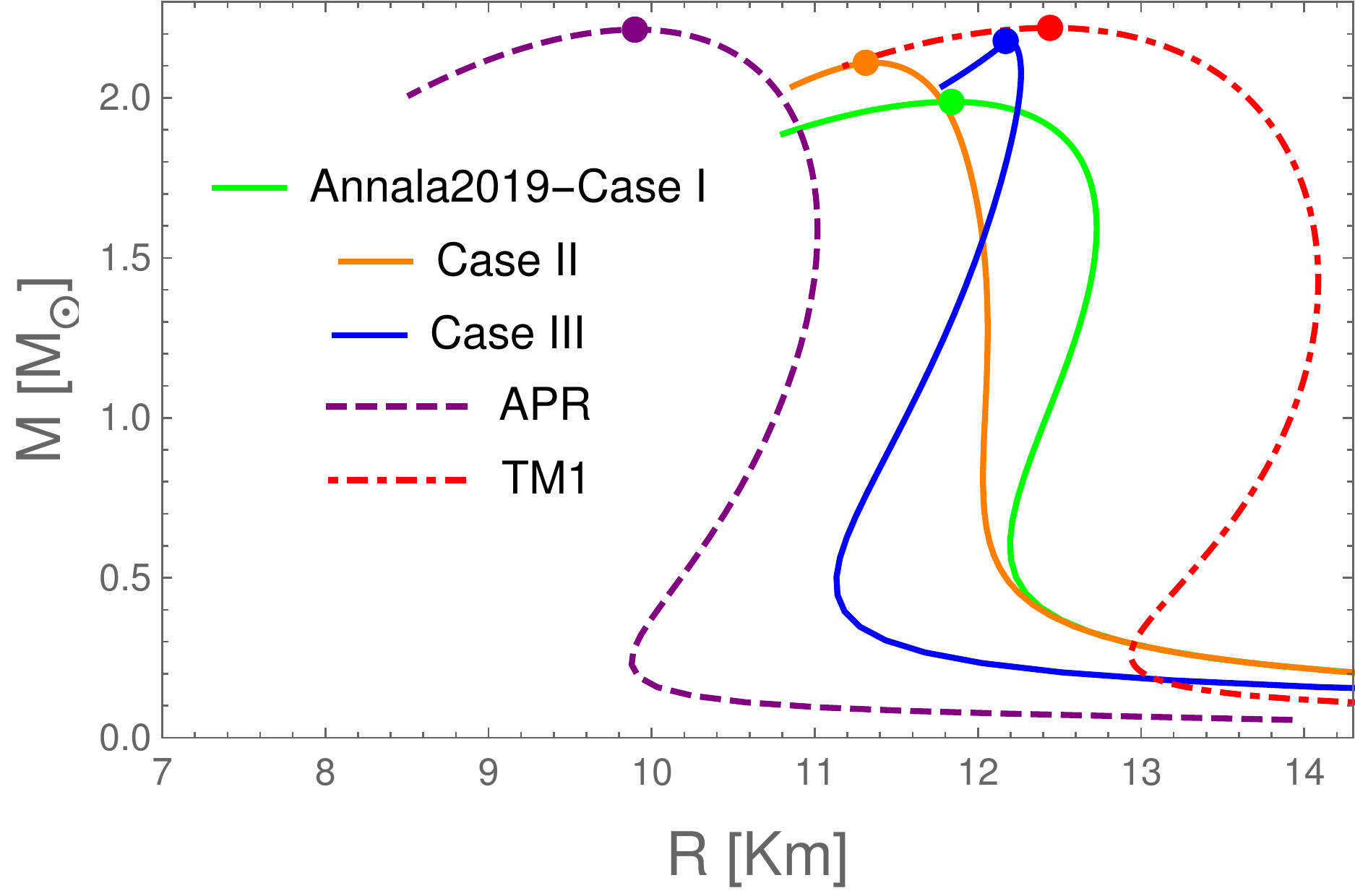}		\includegraphics[width=0.495\textwidth]{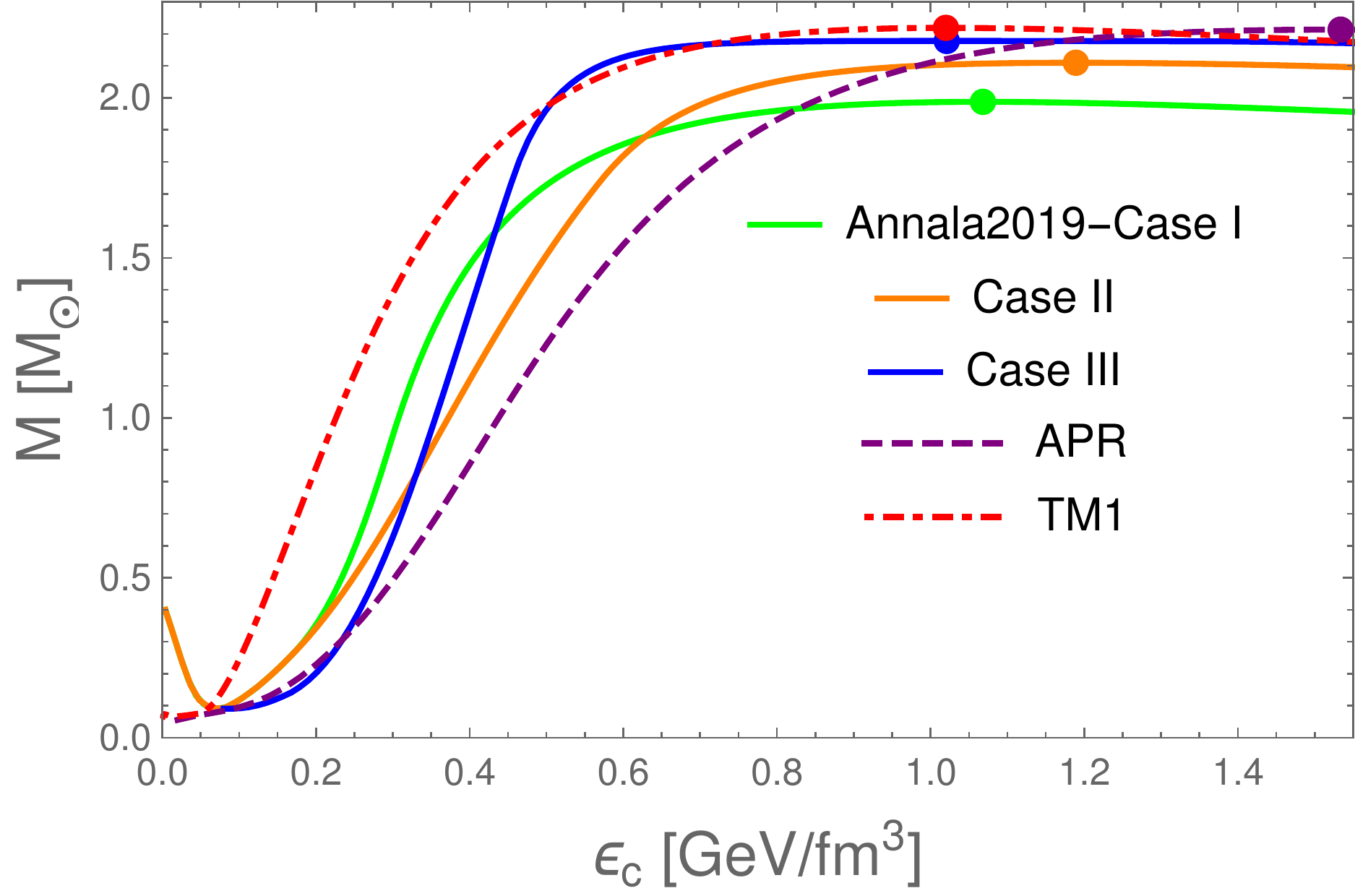}}
\hbox{ \includegraphics[width=0.495\textwidth]{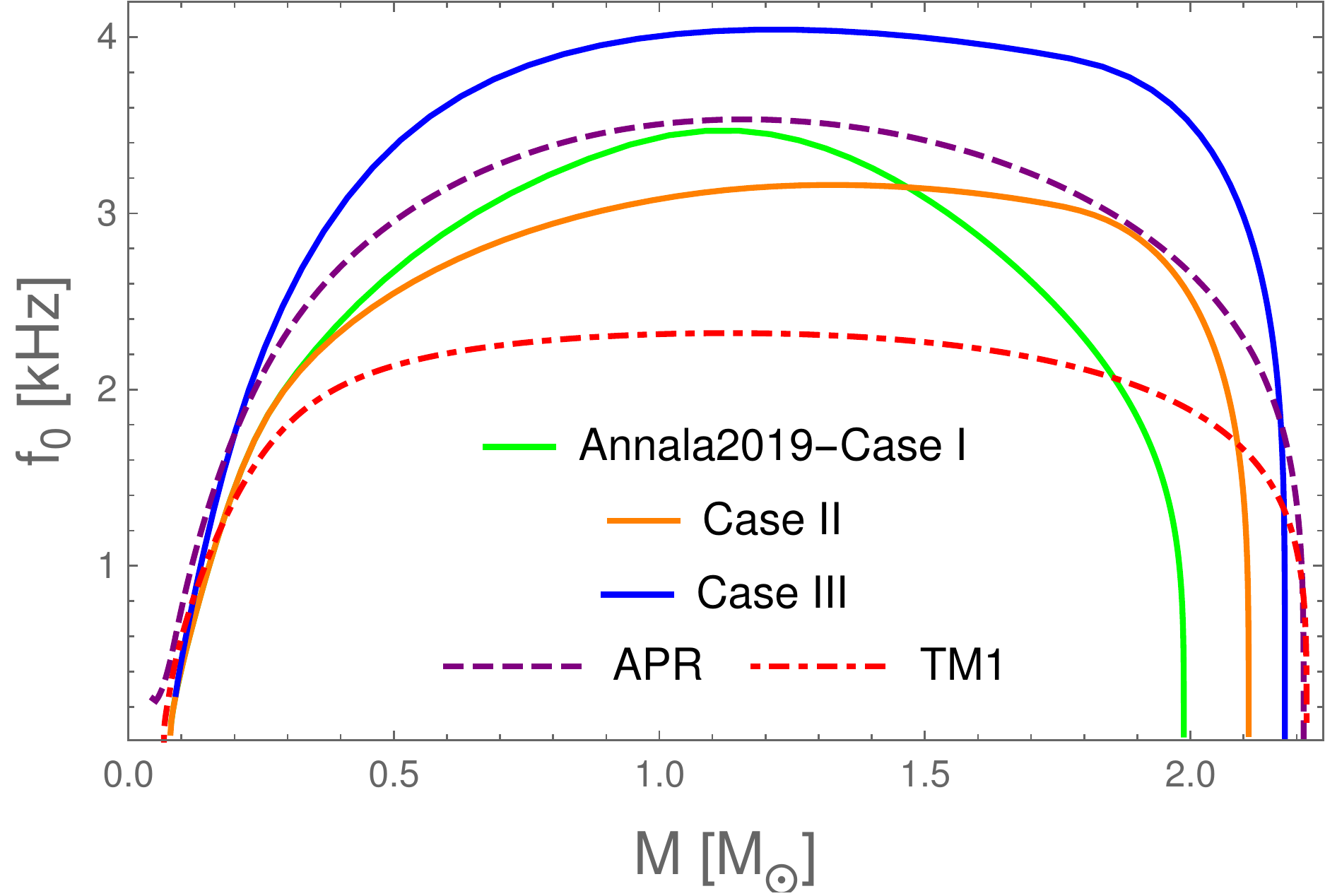}		\includegraphics[width=0.495\textwidth]{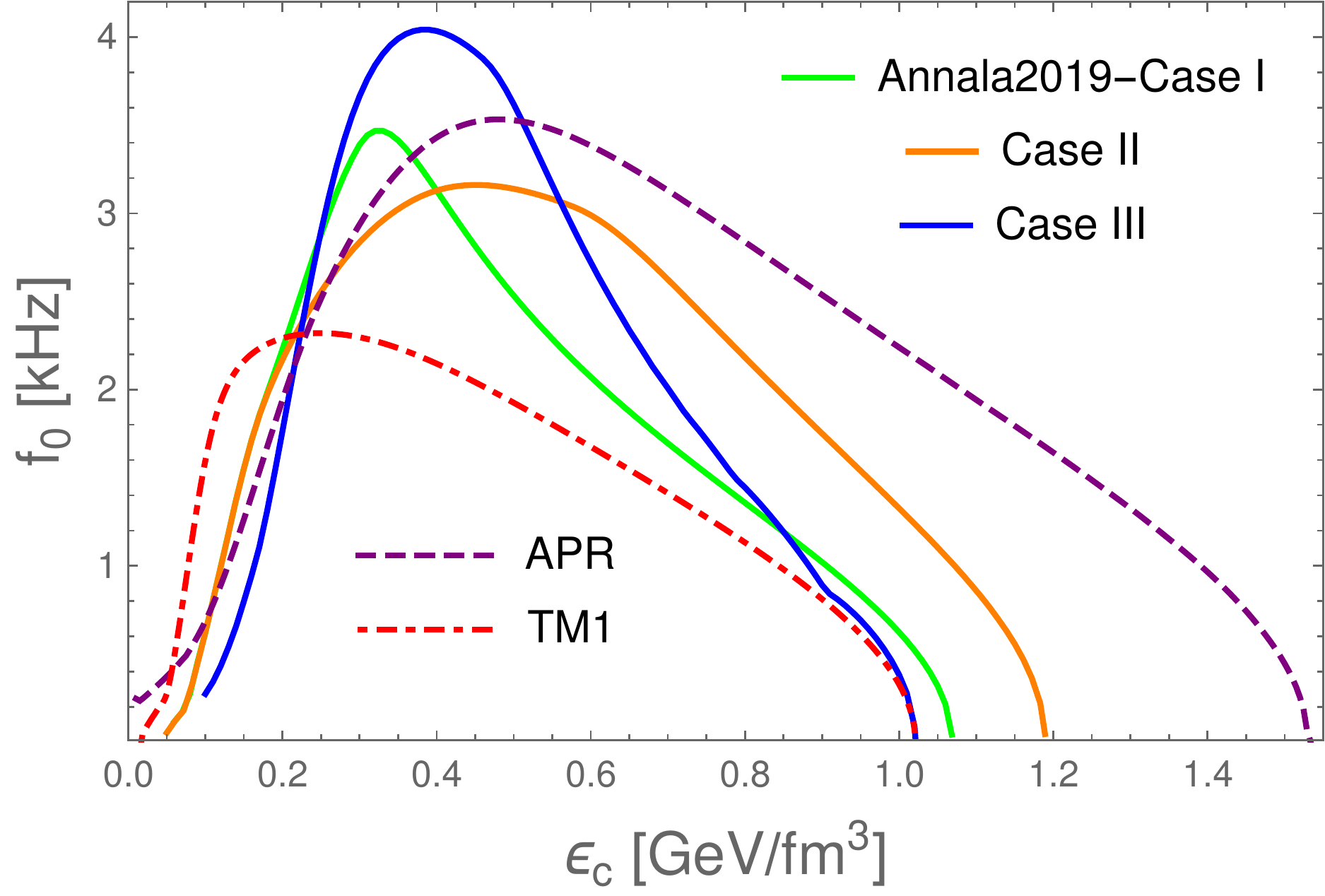}}
\caption{Upper panels: mass-radius diagram and mass as a function of central energy density obtained from Cases I, II and III in Ref. \cite{Annala:2019puf}. Dots indicate the end of the star sequence at $M_{\rm max}$, i.e. $\partial{M}/\partial{\epsilon_{c}}=0$.  Lower panels: corresponding fundamental-mode frequencies ($f_{0}$) as functions of the gravitational mass and central energy density. For comparison we show results from APR \cite{Akmal:1998cf} and TM1 \cite{Shen:1998gq} equations of state for pure nuclear matter.}
\label{fig:Ann2019}
\end{center}
\end{figure*}

From Fig. \ref{fig:Kurk2014} one can see that Case III produces larger values of $f_{0}$ ($\sim 4.5$ kHz) compared to the other cases due to the fast change from being less deformable (more compact) to more deformable (less compact), which is related to a rapid stiffening/softening of $c^{2}_{\rm s, max}=0.95$ when changing polytropes at intermediate densities (see Fig. \ref{fig:EoSsKurk}). This non-standard behavior might be considered unphysical since there is no actual transition that supports this abrupt behavior of the EoS\footnote{Interestingly, nowadays this interpretation is in agreement with the NICER measurements for NS radii with 1.4 $M_{\odot}$ \cite{Miller:2019cac}}. It also goes up with $\epsilon_{c}$ much faster than the standard pure nuclear-matter EoSs even at low densities. On the other hand, Case I matches the behavior of the APR even when both EoSs take very distinct values of $c^{2}_{s}$ at high densities, i.e. being always causal and becoming rapidly acausal, respectively. So, it is hard to distinguish Case I from a hadronic star, at least in the $f_{0}$ vs $\epsilon_{c}$ plane. Case II stands between these extreme behaviors, reaching $c^{2}_{\rm s, max}=0.76$ slowly. Notice that the jumps in $c^{2}_{s}$ do not affect notoriously the associated frequencies.

A final point we can make from these EoSs is that the behavior of $f_{0}$ as a function of $M$ suggests a qualitative distinction between pure and hybrid NS \cite{Gondek:1999ad}: nearly flat $f_{0}(M)$ characterizes nuclear matter cores (such as in the case of APR and TM1 EoSs), whereas QM cores produce very narrow flat regions. For example, Fig. \ref{fig:Kurk2014} shows that Cases I and II behave as hadronic NS and Case III as a hybrid. Nevertheless, as we have already discussed, the last case cannot be considered a realistic situation.

\begin{figure*}[t!]
\begin{center}
\hbox{
\includegraphics[width=0.499\textwidth]{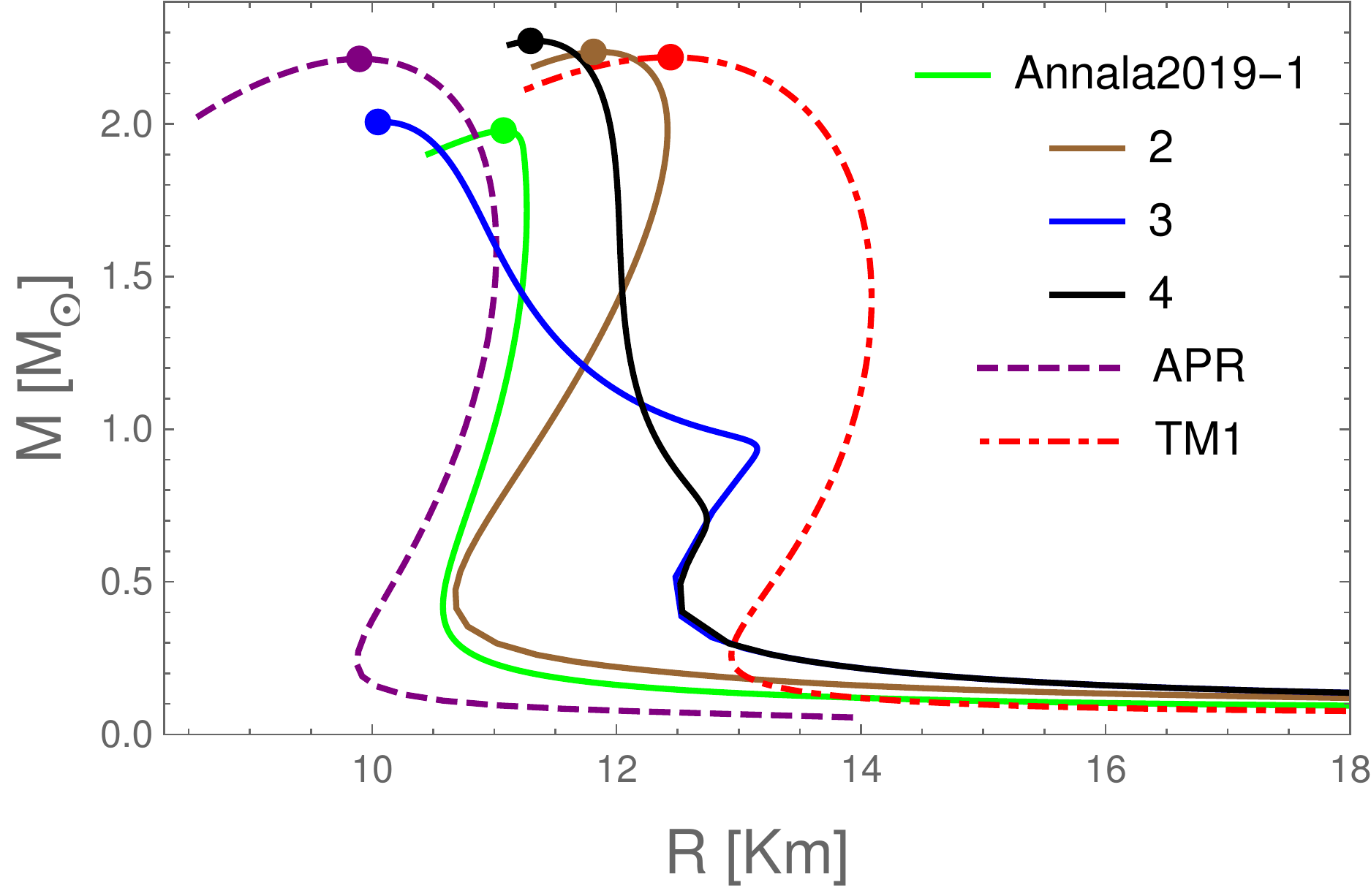}		\includegraphics[width=0.499\textwidth]{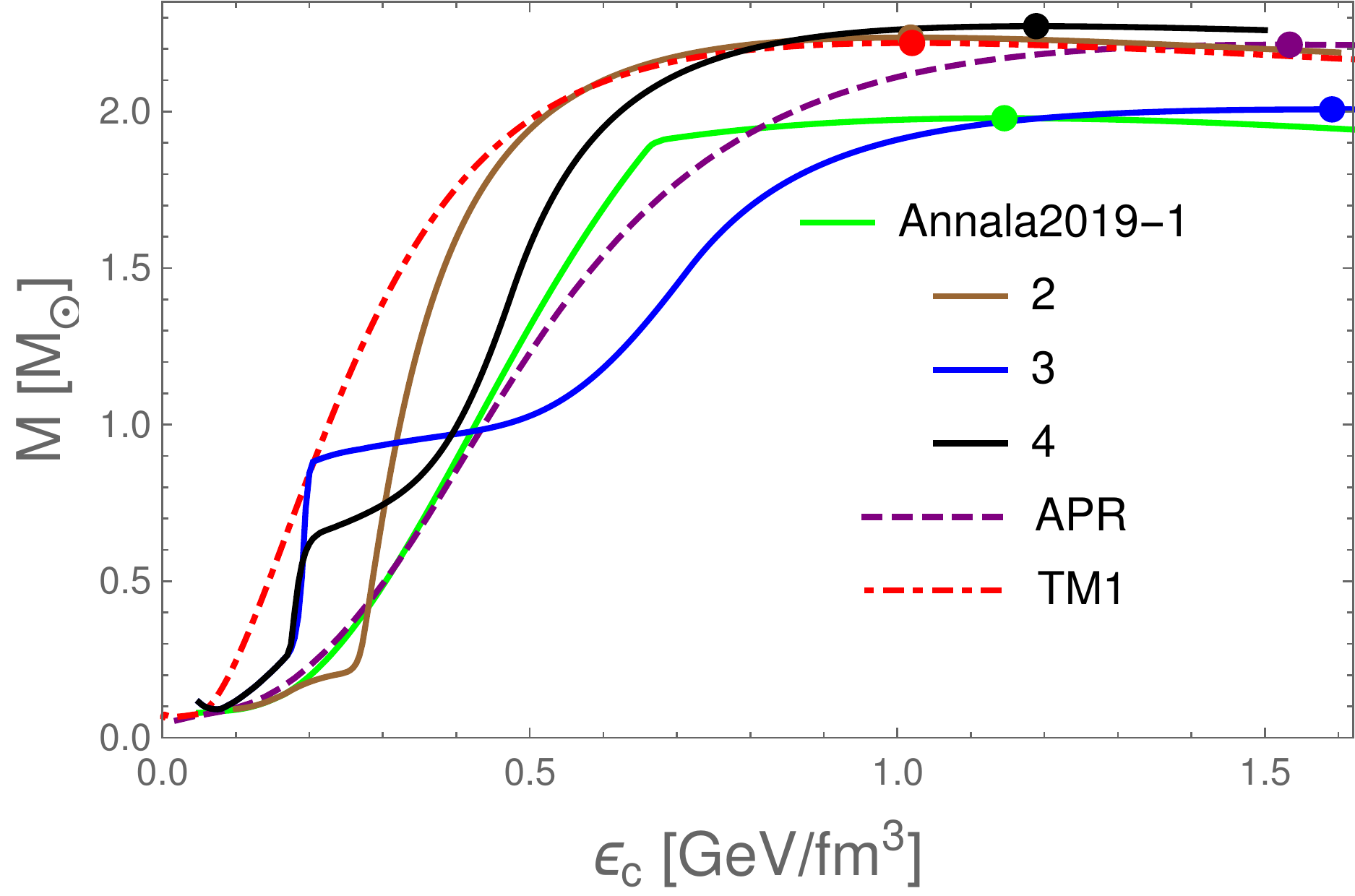}}
\hbox{ \includegraphics[width=0.494\textwidth]{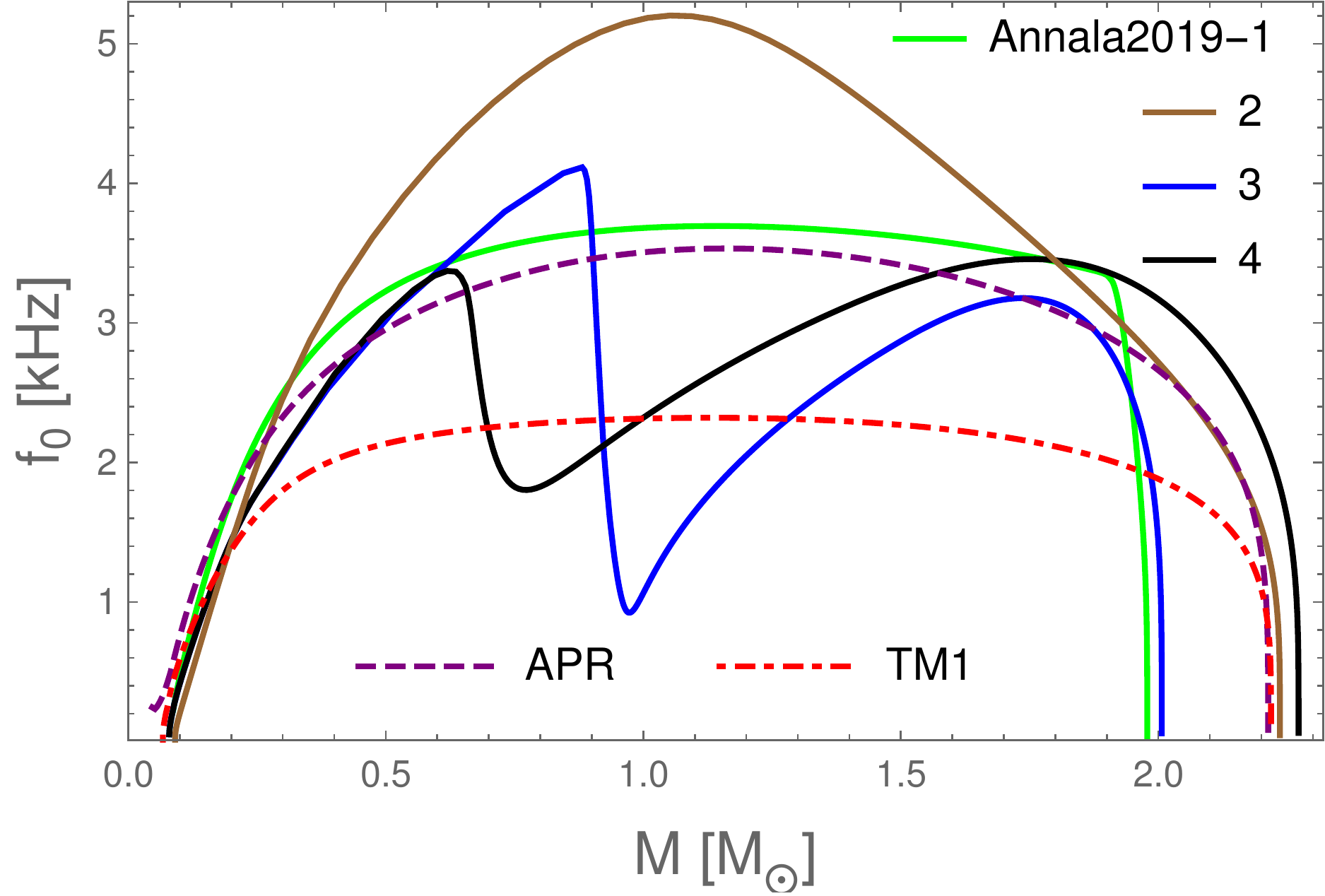}		\includegraphics[width=0.494\textwidth]{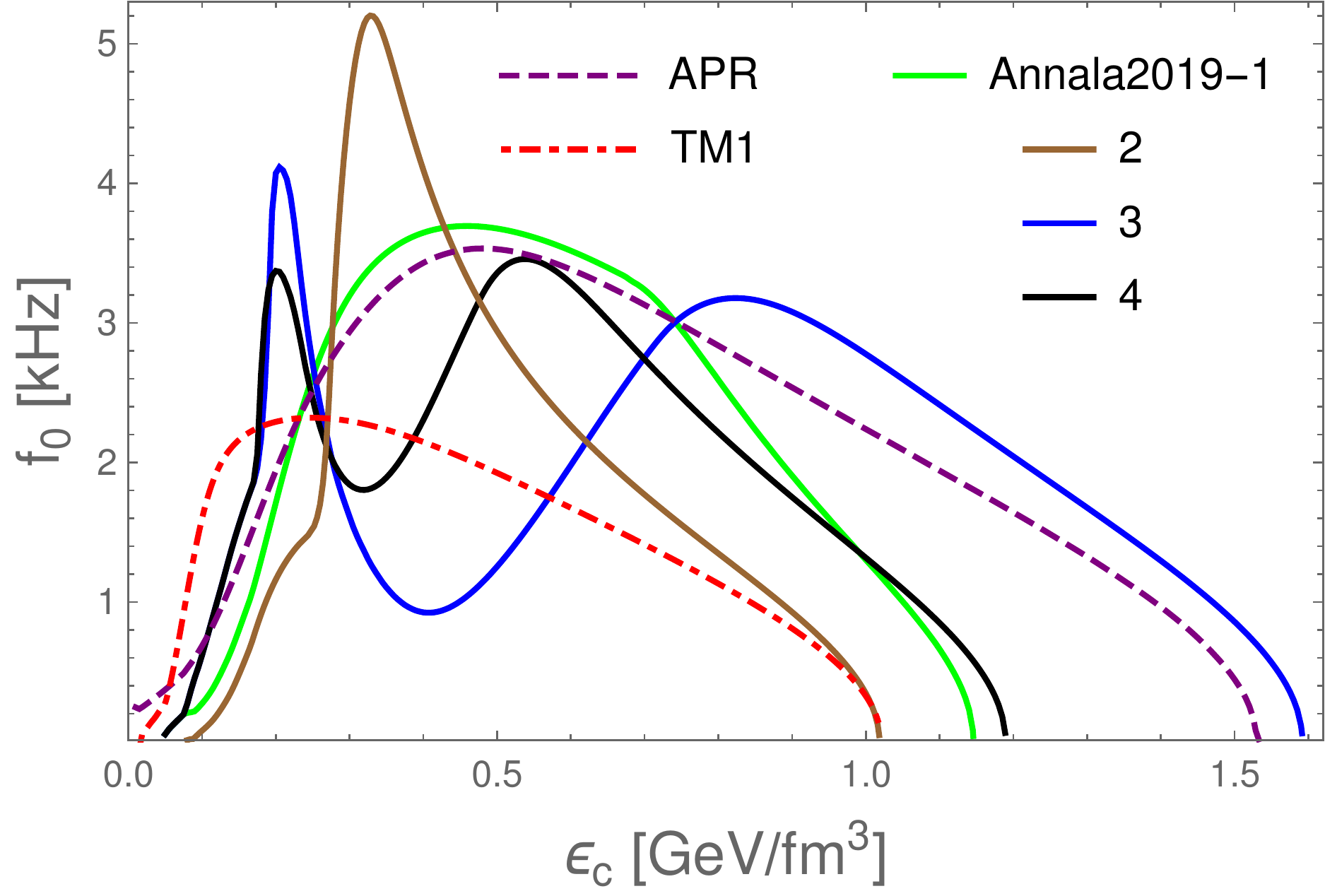}}
\caption{Same as Fig. \ref{fig:Ann2019} now using the EoSs of Annala \textit{et al.} \cite{Annala:2019puf} for Cases 1--4 \cite{Gorda:2019abc}.}
\label{fig:Ann2019x}
\end{center}
\end{figure*}

We now explore the sets of selected EoSs from Ref. \cite{Annala:2019puf}. Recall that, besides the addition of the bounds yielded by the GW170817 event, the authors considered the dependence of the NS matter EoS on $c^{2}_{s}$ (reaching values up to the causality limit) through their speed-of-sound interpolation method. They were motivated by the observation that NS with very high masses and small radii require the EoS to suffer a rapid stiffening somewhere in the core, perhaps due to a hadron-quark transition. In fact, Ref. \cite{Tews:2018kmu} concluded, after a detailed analysis, that $c^{2}_{s}$ should behave non-monotonically at intermediate densities, surpassing the conformal bound and then coming down to approach it again only at very high densities.

In Fig. \ref{fig:EoSs} we show the non-monotonic behavior of the speed of sound as a function of the central energy density. In the first set, Cases I and II display one bump exhibiting a mild breaking of the conformal bound, so containing QM cores\footnote{We note that the aforementioned flatness criterium of Ref. \cite{Gondek:1999ad} to distinguish hybrid NS becomes ambiguous for Case II and only the density profiles can help us solve this issue \cite{Annala:2019puf}.}. Case III, on the other hand, has a second bump at very high densities with much higher $c^{2}_{s, \rm max}$, so yielding purely hadronic stars. Besides, only Case III displays a somewhat notorious kink associated to a fast decrease of $c^{2}_{s}$ before reaching $\epsilon^{\rm max}_{c}$ through a discontinuous transition (see Fig. \ref{fig:EoSs}), after which it becomes unstable and the second bump in $c^{2}_{s}$ begins. This hydrostatic and dynamical equilibrium structure is shown in Fig. \ref{fig:Ann2019}.

As in the case of Kurkela {\it et al.} \cite{Kurkela:2014vha}, changes in compacteness (visible in the M--R diagram) produce a hierarchy of maximal values for the zero-mode frequencies at intermediate densities, from the highest (Case III) to the lowest (Case II) in the $f_{0}$ vs $\epsilon_{c}$ plane. Thus, NS in Cases I and II, having sizeable QM cores, exhibit lower values of $\rm max(f_{0})$. This can be understood by noticing that a large amount of QM dissipates oscillations more rapidly due to its high-density nature\footnote{Of course, these frequencies are smaller (larger) than the APR (TM1) EoS because they reach the conformal limit (see Fig. \ref{fig:EoSs}) faster (more slowly)}. 

The second set of equations of state (Cases 1 to 4) behaves very differently with respect to $c^{2}_{s}$: they exhibit two bumps at relevant NS densities, as shown in Fig. \ref{fig:EoSs}. This is contrast with the analysis of Ref. \cite{Bedaque:2014sqa} in which the authors suggest that the speed of sound could only behave non-monotonically once  (see also Ref. \cite{Pereira:2017rmp} on the effects from phase transitions). It is also remarkable, as shown in Fig. \ref{fig:Ann2019x}, that this set of EoSs (especially Cases 3 and 4) shows a peculiar feature in the mass-radius diagram, which is also manifest in the $M$ vs $\epsilon_{c}$ plot. The effect resembles that of a first-order phase transition, even though the densities are too low for a transition to QM\footnote{The same feature appears in the M--R diagram of Ref. \cite{Capano:2019eae}, where stringent constraints were imposed on the NS radii using information from the GW170817 event plus nuclear theory which best accounts for the density-dependent uncertainties in the EoS.}. This feature is not in line with X-ray observations \cite{Nattila:2017wtj} and brings the question whether the solutions obtained from Cases 1 to 4 would be dynamically stable.

Using the criteria of Ref. \cite{Gondek:1999ad} in Fig. \ref{fig:Ann2019x}, we see that Case 1 behaves as a nucleonic NS, whereas Case 2 as a hybrid NS. This agrees with the criteria of Ref. \cite{Annala:2019puf}, where large $c^{2}_{s}$ favors nuclear matter cores and mild violations of the conformal bound ensure seizeable QM cores. Cases 3 and 4 deserve a more detailed discussion.

The simple approach of verifying the stability by looking at $dM/dR=0$ \cite{Calamai:1970cba} is not adequate to study the non-monotonic Cases 3 and 4. So, we use the full solutions of Gondek's equations, in particular $\xi_{n}$ and $f_{n}$. In this sense, dynamical stability will play the role of a further consistency constraint on the mass-radius diagram. Again, in Fig. \ref{fig:Ann2019x} we see that these cases display, not surprisingly, more pronounced oscillations not found in previous works on hydrid stars, e.g. Refs. \cite{Gupta:2002fk,Pereira:2017rmp}. This lead us to ask whether the associated radial amplitudes, $\xi=\Delta{r}/r$, behave consistently.

A careful analysis of the behavior of $\xi_{n}$ (with $n=0,1,2$) in Cases 3 and 4 indicates that they undergo large (continuous/discontinuous) increments along the {\it interior} of NS, i.e. in the region from the crust's outer layers down to near the core ($r\lesssim{R/2}$), reaching values around $20-40$. In other words, one has radial oscillations with amplitudes much larger than the size of the star. It has been shown that large amplitudes near the core rapidly destabilize the NS due to non-linear general-relativistic effects via, for instance, the breaking of weak equilibrium \cite{Gourgoulhon:1995fde}. These families of stars will most likely undergo a large expansion before collapsing to a black hole. Thus, the corresponding region in the EoS band should be discarded for being dynamically unstable.

The case of pathological amplitudes {\it near the surface}, i.e. within the outer layers of NS, which occurs in Cases 1 and 2, has already been reported in Ref. \cite{Meltzer:1966deo} and is known to be harmless due to its dilute nature. Our calculations show that the same happens for the EoSs of Kurkela {\it et al.} \cite{Kurkela:2014vha} and the first set of Annala {\it et al.} \cite{Annala:2019puf} with amplitudes around $\xi_{n} \lesssim 30$.

\section{Summary and conclusions}
  \label{sec:conclusion}

Using the first-order formalism of Gondek {\it et al. } \cite{Gondek:1997fd}, we have investigated the dynamical stability of hadronic and hybrid NS obtained from a set of constrained equations of state derived from first principles \cite{Kurkela:2014vha,Annala:2019puf}. Using the method of Ref. \cite{Gondek:1999ad}, we showed that it is possible to distinguish between hadronic and hybrid stars from their behavior in $f_{0}$ as a function of $M$ in the set from Ref. \cite{Kurkela:2014vha} and the first set from Ref. \cite{Annala:2019puf} (Cases I, II, III).

The second set of EoSs from Ref. \cite{Annala:2019puf}, especially Cases 3 and 4, behave very differently, though. The NS produced are quasi-stable according to our linear analysis of small oscillations, in the sense that their zero-mode frequency remains positive and real. However, the amplitude of their $\xi$ undergoes a resonance-like behavior in the region near the transition point at the core of NS, reaching values around $\xi \sim 30$, many times larger than the size of the star. Using the results from the full (non-linear) general-relativistic study performed in Ref. \cite{Gourgoulhon:1995fde}, we conclude that these large amplitudes evolve dynamically and make these NS strictly unstable. So, the corresponding region in the EoS band should be discarded. 

The linear stability analysis procedure could also be incorporated systematically in interpolations of the equation of state for NS matter besides requiring that $\partial{M}/\partial{\epsilon}>0$. In future astronomical observations, it might be possible to compare tidal deformabilities of NS with the zero-mode frequencies by mapping their dependence for different NS gravitational masses \cite{Passamonti:2005cz,Passamonti:2007tm}.   
  
\begin{acknowledgments}
The authors thank Tyler Gorda for kindly sharing his constrained equations of state for cold QCD matter and for his useful comments and suggestions. This work was partially supported by INCT-FNA (Process No. 464898/2014-5). 
J. C. J. acknowledges support from FAPESP (Processes No. 2020/07791-7 and No. 2018/24720-6). 
E. S. F. is partially supported by CAPES (Finance Code 001), CNPq, and FAPERJ.
\end{acknowledgments}

\section*{Appendix:\\ Gondek's Stability Framework}

The pair of Lagrangian variables to be solved in the approach of Gondek {\it et al.} \cite{Gondek:1997fd} to small radial pulsation equations for the $n$ mode, $\xi_n$ and $(\Delta{P})_n$, implicitly have a ${e}^{i\omega_n{t}}$ factor and the coefficients entering in Eq. (\ref{Eq.Gondek}) are given by (in units such that $G=c=1$)

\begin{equation*}
\mathcal{Z}(r)=-4{\pi}r{e}^{\lambda}(\epsilon+P)+\frac{1}{\epsilon+P}\frac{dP}{dr},
\end{equation*}

\begin{multline*}
\mathcal{Q}(r,\omega^{2})=-8{\pi}rP(\epsilon+P){e^{\lambda}}+ \\ \frac{r}{\epsilon+P}\left(\frac{dP}{dr}\right)^{2}-4\frac{dP}{dr}+{\omega^{2}r(\epsilon+P)e^{\lambda-\nu}},
\end{multline*}

\begin{equation*}
\mathcal{R}(r)=-\frac{1}{r}\frac{1}{P\Gamma},\hspace{0.5cm} \mathcal{S}(r)=-\frac{1}{\epsilon+P}\frac{dP}{dr}-\frac{3}{r}.
\end{equation*}

Apart from giving us the $P(r)$, $\epsilon(r)$, and $\Gamma(r)$ profiles, solving the TOV equations gives us
the metric profiles for $\nu(r)$ through
  \begin{equation*}
  \frac{d\nu}{dr}=-\frac{2}{P+\epsilon}\frac{dP}{dr},\hspace{0.4cm}\nu(R)=\log\left(1-
  \frac{2M}{R}\right)
  \end{equation*}
and for $\lambda(r)$ (plus the mass profile $\mathcal{M}(r)$ as input)
  \begin{equation*}
  \lambda(r)=-\log\left(1-
  \frac{2\mathcal{M}(r)}{r}\right),\hspace{0.4cm}\lambda(R)=-\nu(R),
  \end{equation*}
both entering Gondek's equations, where we have added their associated boundary conditions \cite{Glendenning:2000}.

The appropriate boundary conditions for Gondek's variables are deduced as follows. For $\Delta{P}$ it is required that $(\Delta{P})_{\rm surface}=0$ since the pressure also goes to zero. At  the center it is found that $\xi_{\rm center}=(\Delta{r}/r)_{\rm center}=1$ by construction since $\Delta{r} \to 0$ as $r \to 0$. We note that for self-consistency of the equations, the solutions for $\xi$ should also be small in order to obtain displacements $\Delta{r}$ smaller compared to the star's size $R$. Furthermore, the requirement of physical smoothness at the NS center and finiteness everywhere in the star gives us $(\Delta{P})_{\rm center}=-3(\xi{P}\Gamma)_{\rm center}$. By following this procedure, we have verified that our code reproduces satisfactorily the mode frequencies of Kokkotas and Ruoff \cite{Kokkotas:2000up} who use a different numerical procedure. 

Finally, the resulting squared frequencies $\omega^{2}_{n}$ should satisfy $\omega^{2}_{0}<\omega^{2}_{1}<\omega^{2}_{2}
<\cdot\cdot\cdot$. Frequencies with $\omega^{2}_{n}>0$ are real with stable and oscillatory modes. On the other hand, if $\omega^{2}_{n}<0$, then the frequency is purely imaginary and the mode is unstable.



\begin{thebibliography}{99}

\bibitem{Glendenning:2000}
  N.~K.~Glendenning,
  {\it Compact Stars -- Nuclear Physics, Particle Physics and General Relativity} (Springer, New York, 2000).

\bibitem{Lattimer:2004pg}
J.~M.~Lattimer and M.~Prakash,
Science \textbf{304}, 536 (2004)

\bibitem{Alford:2006vz}
M.~Alford, D.~Blaschke, A.~Drago, T.~Klahn, G.~Pagliara and J.~Schaffner-Bielich,
Nature {\bf 445}, E7 (2007).

\bibitem{Sagert:2008ka}
I.~Sagert {\it et al.},
Phys.\ Rev.\ Lett.\  {\bf 102}, 081101 (2009).

\bibitem{Weih:2019xvw}
L.~R.~Weih, M.~Hanauske and L.~Rezzolla,
Phys. Rev. Lett. \textbf{124}, 171103 (2020)

\bibitem{Miller:2019cac}
M.~Miller {\it et al.},
Astrophys. J. Lett. \textbf{887}, L24 (2019)

\bibitem{Riley:2019yda}
T.~E.~Riley {\it et al.},
Astrophys. J. Lett. \textbf{887}, L21 (2019)

\bibitem{Demorest:2010bx}
P.~Demorest, T.~Pennucci, S.~Ransom, M.~Roberts and J.~Hessels,
Nature \textbf{467}, 1081 (2010)
  
\bibitem{Antoniadis:2013pzd}
J.~Antoniadis {\it et al.},
Science \textbf{340}, 6131 (2013)
 
\bibitem{Fonseca:2016tux}
E.~Fonseca {\it et al.},
Astrophys. J. \textbf{832}, 167 (2016)

\bibitem{Linares:2018ppq}
M.~Linares, T.~Shahbaz and J.~Casares,
Astrophys. J. \textbf{859}, 54 (2018)

\bibitem{Cromartie:2019kug}
H.~T.~Cromartie {\it et al.},
Nature Astron. \textbf{4}, no.1, 72 (2019)


\bibitem{Tews:2012fj}
I.~Tews, T.~Krüger, K.~Hebeler and A.~Schwenk,
Phys. Rev. Lett. \textbf{110}, 032504 (2013)

\bibitem{Hebeler:2013nza}
K.~Hebeler, J.~Lattimer, C.~Pethick and A.~Schwenk,
Astrophys. J. \textbf{773}, 11 (2013)

\bibitem{Kurkela:2009gj} 
A.~Kurkela, P.~Romatschke and A.~Vuorinen,
Phys.\ Rev.\ D {\bf 81}, 105021 (2010)

\bibitem{Fraga:2013qra} 
E.~S.~Fraga, A.~Kurkela and A.~Vuorinen,
Astrophys.\ J.\  {\bf 781}, L25 (2014)

\bibitem{Kurkela:2014vha} 
A.~Kurkela, E.~S.~Fraga, J.~Schaffner-Bielich and A.~Vuorinen,
Astrophys.\ J.\  {\bf 789}, 127 (2014)

\bibitem{Annala:2017llu} 
E.~Annala, T.~Gorda, A.~Kurkela and A.~Vuorinen,
Phys.\ Rev.\ Lett.\  {\bf 120}, 172703 (2018)

\bibitem{Rezzolla:2017aly}
L.~Rezzolla, E.~R.~Most and L.~R.~Weih,
Astrophys. J. Lett. \textbf{852}, L25 (2018)

\bibitem{Ruiz:2017due}
M.~Ruiz, S.~L.~Shapiro and A.~Tsokaros,
Phys. Rev. D \textbf{97}, 021501 (2018)

\bibitem{TheLIGOScientific:2017qsa}
B.~P.~Abbott \textit{et al.} [LIGO Scientific and Virgo],
Phys. Rev. Lett. \textbf{119}, 161101 (2017)

\bibitem{Annala:2019puf}
E.~Annala, T.~Gorda, A.~Kurkela, J.~N\"{a}ttil\"{a} and A.~Vuorinen,
Nat. Phys. {\bf 16}, 907 (2020)
arXiv:1903.09121 [astro-ph.HE].

\bibitem{Tews:2018iwm}
I.~Tews, J.~Margueron and S.~Reddy,
Phys. Rev. C \textbf{98}, 045804 (2018)

\bibitem{Bazavov:2014pvz}
A.~Bazavov \textit{et al.},
Phys. Rev. D \textbf{90}, 094503 (2014)

\bibitem{Gondek:1997fd}
D.~Gondek, P.~Haensel and J.~Zdunik,
Astron. Astrophys. \textbf{325}, 217 (1997)

\bibitem{Gorda:2019abc}
T. Gorda, private communication.

\bibitem{Bedaque:2014sqa}
P.~Bedaque and A.~W.~Steiner,
Phys. Rev. Lett. \textbf{114}, 031103 (2015)

\bibitem{Gardim:2019xjs}
F.~G.~Gardim, G.~Giacalone, M.~Luzum and J.~Y.~Ollitrault,
Nat. Phys. \textbf{16}, 615 (2020)

\bibitem{Radice:2018ozg}
D.~Radice and L.~Dai,
Eur. Phys. J. A \textbf{55}, 50 (2019)



\bibitem{Gourgoulhon:1995fde}
E.~Gourgoulhon, P.~Haensel and D.~Gondek,
Astron. Astrophys. \textbf{294}, 747 (1995)

\bibitem{Akmal:1998cf}
A.~Akmal, V.~Pandharipande and D.~Ravenhall,
Phys. Rev. C \textbf{58}, 1804 (1998)

\bibitem{Shen:1998gq}
H.~Shen, H.~Toki, K.~Oyamatsu and K.~Sumiyoshi,
Nucl. Phys. A \textbf{637}, 435 (1998)



  

  



  
\bibitem{Gupta:2002fk}
V.~Gupta, V.~Tuli and A.~Goyal,
Astrophys. J. \textbf{579}, 374 (2002)
  
\bibitem{Sahu:2001iv}
P.~Sahu, G.~Burgio and M.~Baldo,
Astrophys. J. Lett. \textbf{566}, L89 (2002)
  
\bibitem{Chandrasekhar:1964zza}
S.~Chandrasekhar,
Phys. Rev. Lett. \textbf{12}, 114 (1964)
  
\bibitem{Haensel:1989wax}
P.~Haensel, J.~L.~Zdunik and R.~Schaeffer,
Astronomy and Astrophysics \textbf{217}, 137 (1989)

\bibitem{Pereira:2017rmp}
J.~P.~Pereira, C.~V.~Flores and G.~Lugones,
Astrophys. J. \textbf{860}, 12 (2018)
  
\bibitem{Haensel:2007}
  P.~Haensel, A.~Y.~Pothekin, and D.~G.~Yakovlev,
  {\it Neutron Stars 1} (Springer-Verlag, New York, 2007).   
  
\bibitem{Gondek:1999ad}
D.~Gondek and J.~Zdunik,
Astron. Astrophys. \textbf{344}, 117 (1999)
  
\bibitem{Tews:2018kmu}
I.~Tews, J.~Carlson, S.~Gandolfi and S.~Reddy,
Astrophys. J. \textbf{860}, 149 (2018)

\bibitem{Capano:2019eae}
C.~D.~Capano, I.~Tews, S.~M.~Brown, B.~Margalit, S.~De, S.~Kumar, D.~A.~Brown, B.~Krishnan and S.~Reddy,
Nature Astron. \textbf{4}, 625 (2020)
  
\bibitem{Nattila:2017wtj}
J.~N\"{a}ttil\"{a}, M.~C.~Miller, A.~W.~Steiner, J.~J.~E.~Kajava, V.~F.~Suleimanov and J.~Poutanen,
Astron. Astrophys. \textbf{608}, A31 (2017)
  
\bibitem{Calamai:1970cba}
G.~Calamai,
Astrophysics and Space Science \textbf{8}, 53 (1970)   
  
\bibitem{Meltzer:1966deo}
D.~W.~Meltzer and K.~S.~Thorne,
Astrophys.\ J.\  {\bf 145}, 514 (1966)    
  
 
\bibitem{Passamonti:2005cz}
A.~Passamonti, M.~Bruni, L.~Gualtieri, A.~Nagar and C.~F.~Sopuerta,
Phys. Rev. D \textbf{73}, 084010 (2006)
 
\bibitem{Passamonti:2007tm}
A.~Passamonti, N.~Stergioulas and A.~Nagar,
Phys. Rev. D \textbf{75}, 084038 (2007)
 
\bibitem{Kokkotas:2000up}
K.~Kokkotas and J.~Ruoff,
Astron. Astrophys. \textbf{366}, 565 (2001)
 
  
 




  



 
   
  
\end{thebibliography}
\end{document}